\documentclass[article]{elsarticle}

\usepackage{lineno,hyperref}
\modulolinenumbers[5]

\journal{Journal of Parallel and Distributed Computing}

\usepackage{amsmath}
\usepackage{relsize}
\usepackage{framed}
\usepackage{graphicx}
\usepackage{caption}
\usepackage[flushleft]{threeparttable}
\usepackage{multirow}
\usepackage{hhline}
\usepackage{rotating}
\usepackage{subcaption}
\usepackage{algpseudocode}
\usepackage{algorithm}
\usepackage{setspace}
\usepackage{url}
\usepackage{soul, color}
\soulregister\cite7
\soulregister\ref7

\algrenewcommand\textproc{}

\begin{document}

\begin{frontmatter}
\title{Parallelized Kendall's Tau Coefficient Computation via SIMD Vectorized Sorting On Many-Integrated-Core Processors}
\fntext[myfootnote]{Preliminary work was presented in the 28th International Symposium on Computer Architecture and High Performance Computing, Los Angeles, USA, 2016}

\author[mymainaddress]{Yongchao Liu\corref{mycorrespondingauthor}}
\ead{yliu@cc.gatech.edu}

\author[mymainaddress]{Tony Pan}
\ead{tpan7@gatech.edu}

\author[mymainaddress]{Oded Green}
\ead{ogreen@gatech.edu}

\author[mymainaddress]{Srinivas Aluru\corref{mycorrespondingauthor}}
\ead{aluru@cc.gatech.edu}

\cortext[mycorrespondingauthor]{Corresponding author}
\address[mymainaddress]{School of Computational Science \& Engineering, Georgia Institute of Technology, Atlanta, GA 30332, USA}

\begin{abstract}
Pairwise association measure is an important operation in data analytics. Kendall's tau coefficient is one widely used correlation coefficient identifying non-linear relationships between ordinal variables. In this paper, we investigated a parallel algorithm accelerating all-pairs Kendall's tau coefficient computation via single instruction multiple data (SIMD) vectorized sorting on Intel Xeon Phis by taking advantage of many processing cores and 512-bit SIMD vector instructions. To facilitate workload balancing and overcome on-chip memory limitation, we proposed a generic framework for symmetric all-pairs computation by building provable bijective functions between job identifier and coordinate space. Performance evaluation demonstrated that our algorithm on one 5110P Phi achieves two orders-of-magnitude speedups over 16-threaded MATLAB and three orders-of-magnitude speedups over sequential R, both running on high-end CPUs. Besides, our algorithm exhibited rather good distributed computing scalability with respect to number of Phis. Source code and datasets are publicly available at \url{http://lightpcc.sourceforge.net}.
\end{abstract}

\end{frontmatter}


\section{Introduction}
Identifying interesting pairwise association between variables is an important operation in data analytics. In bioinformatics and computational biology, one typical application is to mine gene co-expression relationship via gene expression data, which can be realized by query-based gene expression database search \cite{zhu2015targeted} or gene co-expression network analysis \cite{steuer2002mutual}. For gene expression database search, it targets to select the subject genes in the database that are co-expressed with the query gene. One approach is to first define some pairwise correlation/dependence measure over gene expression profiles across multiple samples (gene expression profiles for short) and then rank query-subject gene pairs by their scores. For gene co-expression networks, nodes usually correspond to genes and edges represent significant gene interactions inferred from the association of gene expression profiles. To construct a gene co-expression network, all-pairs computation over gene expression profiles is frequently conducted based on linear (e.g. \cite{butte1999unsupervised} \cite{mutwil2011planet} \cite{gobbi2015null}) or non-linear (e.g. \cite{margolin2006aracne} \cite{aluru2012reverse} \cite{lachmann2016aracne}) co-expression measures. A variety of correlation/dependence measures have been proposed in the literature and among them, Pearson's product-moment correlation coefficient \cite{lee1988thirteen} (or Pearson's $r$ correlation) is the most widely used correlation measure \cite{song2012comparison}. However, this correlation coefficient is only applicable to linear correlations. In contrast, Spearman's rank correlation coefficient \cite{spearman1904spearman} (or Spearman's $\rho$ coefficient) and Kendall's rank correlation coefficient \cite{kendall1948rank} (or Kendall's $\tau$ coefficient) are two commonly used measures for non-linear correlations \cite{wang2015efficient}. Spearman's $\rho$ coefficient is based on Pearson's $r$ coefficient but applies to ranked variables, while Kendall's $\tau$ coefficient tests the association between ordinal variables. These two rank-based coefficients were shown to play complementary roles in the cases when Pearson's $r$ is not effective \cite{xu2013comparative}. Among other non-linear measures, mutual information \cite{darbellay1999estimation} \cite{daub2004estimating} \cite{liang2008gene}, Euclidean distance correlation \cite{szekely2007measuring} \cite{kosorok2009brownian}, Hilbert-Schmidt information criterion \cite{gretton2005measuring}, and maximal information criterion \cite{reshef2011detecting} are frequently used as well. In addition, some unified frameworks for pairwise dependence assessments were proposed in the literature (e.g. \cite{song2012feature}).

Kendall's $\tau$ coefficient ($\tau$ coefficient for short) measures the ordinal correlation between two vectors of ordinal variables. Given two ordinal vectors $u=\{u_1, u_2, ..., u_n\}$ and $v=\{v_1, v_2, ..., v_n\}$, where variables $u_i$ and $v_i$ are both ordinal ($0\leq i < n$), the $\tau$ coefficient computes the correlation by counting the number of concordant pairs $n_c$ and the number of discordant pairs $n_d$ by treating $u_i$ and $v_i$ as a joint ordinal variable $(u_i, v_i)$. For the $\tau$ coefficient, a pair of observations $(u_i, v_i)$ and $(u_j, v_j)$, where $i\neq j$, is deemed as concordant if $u_i > u_j$ and $v_i > v_j$ or $u_i < u_j$ and $v_i < v_j$, and discordant if $u_i > u_j$ and $v_i < v_j$ or $u_i < u_j$ and $v_i > v_j$. Note that if $u_i = u_j$ or $v_i = v_j$, this pair is considered neither concordant nor discordant.

In our study, we will consider two categories of $\tau$ coefficient, namely Tau-a (denoted as $\tau_A$) and Tau-b (denoted as $\tau_B$). $\tau_A$ does not take into account tied elements in each vector and is defined as
\begin{equation}
\tau_A = \frac{n_c - n_d}{n_0}
\label{equation:tau_a}
\end{equation}
where $n_0=n(n-1)/2$. If all elements in each vector are distinct, we have $n_c + n_d = n_0$ and can therefore re-write Equation (\ref{equation:tau_a}) as
\begin{equation}
\tau_A = \frac{n_0 - 2n_d}{n_0} = 1 - \frac{2n_d}{n_0}
\label{equation:tau_a_alias}
\end{equation}
As opposed to $\tau_A$, $\tau_B$ makes adjustments for ties and is computed as
\begin{equation}
\tau_B = \frac{n_c - n_d}{\sqrt{(n_0 - n_1)(n_0 - n_2)}}
\label{equation:tau_b}
\end{equation}
where $n_1 = \sum_i{u'_i(u'_i - 1) / 2}$ and $n_2=\sum_i{v'_i(v'_i - 1) / 2}$. $u'_i$ ($v'_i$) denotes the cardinality of the $i$-th group of ties for vector $u$ ($v$). Within a vector, each distinct value defines one tie group and this value acts as the identifier of the corresponding group. The cardinality of a tie group is equal to the number of elements in the vector whose values are identical to the identifier of the tie group. If all elements in either vector are distinct, $\tau_B$ will equal $\tau_A$ as $n_1 = n_2 = 0$. This indicates $\tau_A$ is a special case of $\tau_B$ and an implementation of $\tau_B$ will cover $\tau_A$ inherently. In addition, from the definitions of $\tau_A$ and $\tau_B$, we can see that the computation of the $\tau$ coefficient between $u$ and $v$ is commutable.

Besides the values of correlation coefficients, some applications need to calculate $P$-value statistics to infer statistical significance between variables. For this purpose, one approach is permutation test \cite{chang2016gsa}. However, a permutation test may need a substantial number of pairwise $\tau$ coefficient computation \cite{abdi2007kendall} even for moderately large $n$, thus resulting in prohibitively long times for sequential execution. In the literature, parallelizing pairwise $\tau$ coefficient computation has not yet been intensively explored. One recent work is from Wang \textit{et al}. \cite{wang2014optimising}, which accelerated the sequential $\tau$ coefficient computation in R \cite{team2013r} based on Hadoop MapReduce \cite{dean2008mapreduce} parallel programming model. In \cite{wang2014optimising}, the sequential all-pairs $\tau$ coefficient implementation in R was shown extremely slow on large-scale datasets. In our study, we further confirmed this observation through our performance assessment (refer to section \ref{sec:matlab_r}).

In this paper, we parallelized all-pairs $\tau$ coefficient computation on Intel Xeon Phis based on Many-Integrated-Core (MIC) architecture, the first work accelerating all-pairs $\tau$ coefficient computation on MIC processors to the best of our knowledge. This work is a continuation from our previous parallelization of all-pairs Pearson's $r$ coefficient on Phi clusters \cite{liu2016parallel} and further enriches our LightPCC library (\url{http://lightpcc.sourceforge.net}) targeting parallel pairwise association measures between variables in big data analytics. In this work, we have investigated three variants, namely the na\"{i}ve variant, the generic sorting-enabled (GSE) variant and the vectorized sorting-enabled (VSE) variant, built upon three pairwise $\tau$ coefficient kernels, i.e. the na\"{i}ve kernel, the GSE kernel and the VSE kernel, respectively. Given two ordinal vectors $u$ and $v$ of $n$ elements each, the na\"{i}ve kernel enumerates all possible pairs of joint variables $(u_i, v_i)$ ($0\leq i < n$) to obtain $n_c$ and $n_d$, resulting in $O(n^2)$ time complexity. In contrast, both the GSE and VSE kernels take sorting as the core and manage to reduce the time complexity to $O(n\log n)$.

Given $m$ vectors of $n$ elements each, the overall time complexity would be $O(m^2n^2)$ for the na\"{i}ve variant and $O(m^2n\log n)$ for the GSE and VSE variants. The VSE variant enhances the GSE one by exploiting 512-bit wide single instruction multiple data (SIMD) vector instructions in MIC processors to implement fast SIMD vectorized pairwise merge of sorted subarrays. Furthermore, to facilitate workload balancing and overcome on-chip memory limitation, we investigated a generic framework for symmetric all-pairs computation by pioneering to build a provable, reversible and bijective relationship between job identifier and coordinate space in a job matrix. 

The performance of our algorithm was assessed using a collection of real whole human genome gene expression datasets. Our experimental results demonstrates that the VSE variant performs best on both the multi-threaded CPU and Phi systems, compared to the other two variants. We further compared our algorithm with the all-pairs $\tau$ coefficient implementations in the widely used MATLAB \cite{matlab2015b} and R \cite{team2013r}, revealing that our algorithm on a single 5110P Phi achieves up to 812 speedups over 16-threaded MATLAB and up to 1,166 speedups over sequential R, both of which were benchmarked on high-end CPUs. In addition, our algorithm exhibited rather good distributed computing scalability with respect to number of Phis.
\section{Intel Many-Integrated-Core (MIC) Architecture}
Intel MIC architecture targets to combine many Intel processor cores onto a single chip and has already led to the release of two generations of MIC processors. The first generation is code named as Knights Corner (KNC) and the second generation code named as Knights Landing (KNL) \cite{sodani2016knights}. KNC is a PCI Express (PCIe) connected coprocessor that must be paired with Intel Xeon CPUs. KNC is actually a shared-memory computer \cite{jeffers2013intel} with full cache coherency over the entire chip and running a specialized Linux operating system over many cores. Each core adopts an in-order micro-architecture and has four hardware threads offering four-way simultaneous multithreading. Besides scalar processing, each core is capable of vectorized processing from a newly designed vector processing unit (VPU) featuring 512-bit wide SIMD instruction set architecture (ISA). For KNC, each core has only one VPU and this VPU is shared by all active hardware threads running on the same core. Each 512-bit vector register can be split to either 8 lanes with 64 bits each or 16 lanes with 32 bits each. Note that the VPU does not support legacy SIMD ISAs such as the Streaming SIMD extensions (SSE) series. As for caches, each core has separate L1 instruction and data caches of size 32 KB each, and a 512 KB L2 cache interconnected via a bidirectional ring bus with the L2 caches of all other cores to form a unified shared L2 cache over the chip. The cache line size is 64 bytes. In addition, two usage models can be used to invoke KNC: offload model and native model, where the former relies on compiler pragmas/directives to offload highly-parallel parts of an application to KNC, while the latter treats KNC as symmetric multiprocessing computers. While primarily focusing on KNC Phis in this work, we note that our KNC-based implementations can be easily ported onto KNL processors, as KNL implements a superset of KNC instruction sets. We expect that our implementations will be portable to future Phis as well.
\section{Pairwise Correlation Coefficient Kernels}
For the $\tau$ coefficient, we have investigated three pairwise $\tau$ coefficient kernels: the na\"{i}ve kernel, the GSE kernel and the VSE kernel. From its definition, it can be seen that the Kendall's $\tau$ coefficient only depends on the order of variable pairs. Hence, given two ordinal vectors, we can first order all elements in each vector, then replace the original value of every element with its rank in each vector, and finally conduct the $\tau$ coefficient computation on the rank transformed new vectors. This rank transformation does not affect the resulting coefficient value, but could streamline the computation, especially for ordinal variables in complex forms of representation. Moreover, this transformation needs to be done only once beforehand for each vector. Hence, we will assume that all ordinal vectors have already been rank transformed in the following discussions. For the convenience of discussion, Table \ref{tab:notation} shows a list of notions used across our study.
\begin{table}
\centering
\caption{A list of notions used}
\label{tab:notation}
\begin{tabular}{llll}
\hline
Notion&	Description&\\
\hline
$m$&	number of vectors\\
$n$&	number of elements per vector\\
$u$&	vector $u$ of $n$ elements, likewise for $v$\\
$u_i$&	$i$-th element of vector $u$, likewise for $v_i$\\
$n_c$&	number of concordant variable pairs\\
$n_d$&	number of discordant variable pairs\\
$n_0$& 	$n(n-1)/2$\\
$n_1$& $\sum_i{u'_i(u'_i - 1) / 2}$, $u'_i$ is size of the $i$-th tie group in $u$\\
$n_2$& $\sum_i{v'_i(v'_i - 1) / 2}$, $v'_i$ is size of the $i$-th tie group in $v$\\
$n_3$&	$\sum_i{w_i(w_i - 1) / 2}$, $w_i$ is size of the $i$-th joint tie group for $u$ and $v$\\
$\tau_A$&	$(n_c - n_d)/n_0$\\
$\tau_B$&	$(n_c - n_d)/\sqrt{(n_0 - n_1)(n_0 - n_2)}$\\
$S_X$&		a sorted subarray\\
$|S_X|$&	length of the sorted subarray $S_X$\\
$vX$&		a 512-bit SIMD vector with 16 32-bit integer lanes\\
\hline
\end{tabular}
\end{table}
\subsection{Na\"{i}ve Kernel}
The na\"{i}ve kernel enumerates all possible combinations of joint variables $(u_i, v_i$) ($0\leq i < n $) and counts the number of concordant pairs $n_c$ as well as the number of discordant pairs $n_d$. As mentioned above, given two joint variable pairs $(u_i, v_i)$ and $(u_j, v_j$) ($i\neq j$), they are considered concordant if $u_i > u_j$ and $v_i > v_j$ or $u_i <	 u_j$ and $v_i < v_j$, discordant if $u_i > u_j$ and $v_i < v_j$ or $u_i < u_j$ and $v_i > v_j$, and neither concordant nor discordant if $u_i = u_j$ or $v_i = v_j$. Herein, we can observe that the two joint variables are concordant if and only if the value of $(u_i - u_j)\times (v_i - v_j)$ is positive; discordant if and only if the value of $(u_i - u_j)\times (v_i - v_j)$ is negative; and neither concordant nor discordant if and only if the value of $(u_i - u_j)\times (v_i - v_j)$ is equal to zero. In this case, in order to avoid branching in execution paths (particularly important for processors without hardware branch prediction units), we compute the value of $n_c - n_d$ by examining the sign bit of the product of $u_i - u_j$ and $v_i - v_j$ (refer to lines 2 and 11 in Algorithm \ref{alg:naive_kernel}).
\begin{algorithm}
\caption{Pseudocode of our na\"{i}ve kernel}
\label{alg:naive_kernel}
\fontsize{7.5}{7.75}\selectfont
\begin{algorithmic}[1]
\Function{calc\_sign}{$v$}
	\State \Return {$(v > 0) - (v < 0)$;} \Comment{return 1 if $v > 0$, -1 if $v < 0$ and 0, otherwise}
\EndFunction

\vspace{1em}	

\Function{kendall\_tau\_a\_na\"{i}ve}{$u$, $v$, $n$}

	\State{$norminator$ = 0;} \Comment{$norminator$ represents $n_c - n_d$}
    \For{\textit{i} = 1; $i < n$; ++$i$}
    	\State{$a = u_i$; $b = v_i$;}
    	\State{\#pragma vector aligned}
		\State{\#pragma simd reduction(+:$nominator$)}
    	\For{$j$ = 0; $j < i$; ++$j$}
        	\State{$nominator$ += {\tt calc\_sign}($(a - u_j))\times (b - v_j)$);}         	\Comment{compute $n_c - n_d$}
        \EndFor
    \EndFor

    \State \Return {$\frac{nominator}{n(n-1)/2}$;}
\EndFunction
\end{algorithmic}
\end{algorithm}

Algorithm \ref{alg:naive_kernel} shows the pseudocode of the na\"{i}ve kernel. From the code, the na\"{i}ve kernel has a quadratic time complexity in a function of $n$, but its runtime is independent of the actual content of $u$ and $v$, due to the use of function {\tt calc\_sign}. Meanwhile, the space complexity is $O(1)$. Note that this na\"{i}ve kernel is only used to compute $\tau_A$.
\subsection{Generic Sorting-enabled Kernel}
Considering the close relationship between calculating $\tau$ and ordering a list of variables, Knight \cite{knight1966computer} proposed a merge-sort-like divide-and-conquer approach with $O(n\log n)$ time complexity, based on the assumption that no element tie exists within any vector. As this assumption is not always the case, Knight did mention this drawback and suggested an approximation method by averaging counts, rather than propose an exact solution. In this subsection, we investigate an exact sorting-enabled solution to address both cases: with or without element ties within any vector, together in a unified manner.

Given two ordinal vectors $u$ and $v$, this GSE kernel generally works in the following five steps.
\begin{itemize}
\item Step1 sorts the list of joint variables $(u_i, v_i)$ ($0\leq i < n$) in ascending order, where the joint variables are sorted first by the first element $u_i$ and secondarily by the second element $v_i$. In this step, we used quicksort via the standard {\tt qsort} library routine, resulting in $O(n\log n)$ time complexity.
\item Step2 performs a linear-time scan over the sorted list to compute $n_1$ (refer to Equation (\ref{equation:tau_b})) by counting the number of groups consisting of tied values as well as the number of tied values in each group. Meanwhile, we compute a new value $n_3$ for joint ties, with respect to the pair ($u_i$, $v_i$), as $\sum_i w_i(w_i-1)/2$, where $w_i$ represents the number of jointly tied values in the $i$-th group of joint ties for $u$ and $v$.
\item Step3 counts the number of discordant pairs $n_d$ by re-sorting the sorted list obtained in Step1 in ascending order of all elements in $v$ via a merge sort procedure that can additionally accumulate the number of discordant joint variable pairs each time two adjacent sorted subarrays are merged. The rationale is as follows. Firstly, when merging two adjacent sorted subarrays, we count the number of discordant pairs by only performing pairwise comparison between joint variables from distinct subarrays. In this way, we can ensure that every pair of joint variables will be enumerated once and only once during the whole Step3 execution. Secondly, given two adjacent sorted subarrays to merge, it is guaranteed that the first value (corresponding to $u$) of every joint variable in the left subarray (with the smaller indices) is absolutely less than or equal to the first value of every joint variable in the right subarray (with the larger indices), due to the sort conducted in Step1. In particular, when the first value is identical for the two joint variables from distinct subarrays, the second value (corresponding to $v$) of the joint variable from the left subarray is also absolutely less than or equal to the second value of the joint variable from the left subarray. This means that discordance occurs only if the second value of a joint variable from the right subarray is less than the second value of a joint variable from the left subarray. Therefore, the value of $n_d$ can be gained by accumulating the number of occurrences of the aforementioned discordance in every pairwise merge of subarrays. Algorithm \ref{alg:gse_kernel} shows the pseudocode of counting discordant pairs with out-of-place pairwise merge of adjacent sorted subarrays, where the time complexity is $O(n\log n)$ and the space complexity is $O(n)$.
\item Step4 performs a linear-time scan over the sorted list obtained in Step3 to compute $n_2$ (see Equation (\ref{equation:tau_b})) in a similar way to Step2. This works because the sorted list actually corresponds to a sorted list of all elements in $v$.
\item Step 5 computes the numerator $n_c - n_d$ in Equations (\ref{equation:tau_a}) and (\ref{equation:tau_b}) as $n_0 - n_1 - n_2 + n_3 - 2n_d$. Note that if there is no tie in each vector, $n_1$, $n_2$ and $n_3$ will all be zero. In this case, $n_c - n_d$ will be equal to $n_0 - 2n_d$ as shown in Equation (\ref{equation:tau_a_alias}).
\end{itemize}

From the above workflow, we can see that the GSE kernel takes into account tied elements within each vector. Unlike the na\"{i}ve kernel that computes $\tau_A$, the GSE kernel targets the computation of $\tau_B$. As mentioned above, $\tau_A$ is actually a special case of $\tau_B$ and an algorithm for $\tau_B$ will inherently cover $\tau_A$. Therefore, our GSE kernel is able to calculate both $\tau_A$ and $\tau_B$ in a unified manner. In addition, the GSE kernel has $O(n\log n)$ time complexity, since the time complexity is $O(n\log n)$ for both Step1 and Step3, $O(n)$ for both Step2 and Step4, and $O(1)$ for Step 5.
\begin{algorithm}
\caption{Pseudocode of Step3 of our GSE kernel}
\label{alg:gse_kernel}
\fontsize{7.5}{7.25}\selectfont

\begin{algorithmic}[1]
\Function{gse\_merge}{$in$, $out$, $left$, $mid$, $right$}
	\State{$l = p = left$; $r = mid$; $n_d$ = 0;}
    
    \While{$l < mid$ \&\& $r < right$}     \Comment{merge two sorted subarrays}
    	\If{$in[r].v < in[l].v$}
        
        	\State{$n_d += mid - l$;} \Comment{count discordant pairs}
            \State{out[p++] = in[r++];}
        \Else
        	\State{out[p++] = in[l++];}
        \EndIf
    \EndWhile
    
    \State \Return {$n_d$;}
\EndFunction

\vspace{1em}

\Function{kendall\_tau\_b\_step3\_gse}{$pairs$, $buffer$, $n$}
	\State{$n_d = 0; in = pairs; out = buffer;$}
    \For{$s$ = 1;  $s < n$; $s$ *= 2}
	    \For{$l$ = 0; $l <n$; $l$ += 2 * $s$}
        	\State{$m = \min(l + s, n)$;}
            \State{$r = \min(l + 2 * s, n)$;}
            
            \State{$n_d$ += {\tt gse\_merge}($in$, $out$, $l$, $m$, $r$);} \Comment{Perform out-of-place merge}
        \EndFor
        \State{{\tt swap}($in$, $out$);} \Comment{swap in and out}
  	\EndFor
    \If{pairs != in}
    	\State{{\tt memcpy}($pairs, in, n$ * sizeof(*$pairs$));}
    \EndIf
    
    \State \Return {$n_d$;}
\EndFunction
\end{algorithmic}
\end{algorithm}
\subsection{Vectorized Sorting-enabled Kernel}
The VSE kernel enhances the GSE kernel by employing 512-bit SIMD vector instructions on MIC processors to implement vectorized pairwise merge of sorted subarrays. In contrast with the GSE kernel, the VSE kernel has made the following algorithmic changes. The first is packing a rank variable pair $(u_i, v_i$) into a signed 32-bit integer. In this packed format, each variable is represented by 15 bits in the 32-bit integer, with $u_i$ taking the most significant 16 bits and $v_i$ the least significant 16 bits. In this case, the VSE kernel limits the maximum allowable vector size $n$ to $2^{15}-1=32,767$. The second is that due to packing, we replace the generic variable pair quicksort in Step1 with an integer sorting method, and re-implement the discordant pair counting algorithm in Step3 (see Algorithm \ref{alg:gse_kernel}) based on integer representation. In Step1, sorting packed integers is equivalent to sorting generic variable pairs $(u_i, v_i)$, since within any packed integer $u_i$ sits in higher bits and $v_i$ in lower bits. In Step3, an additional preprocessing procedure is needed to reset to zero the most significant 16 bits of each packed integer (corresponding to $u$). In our implementation, we split a 512-bit SIMD vector into a 16-lane 32-bit-integer vector and then investigated a vectorized merge sort for Step1 and a vectorized discordant pair counting algorithm for Step3, both of which use the same pairwise merge method and also follow a very similar procedure. Algorithm \ref{alg:vse_kernel} shows the pseudocode of Step3 of our VSE kernel.
\begin{algorithm}
\centering
\caption{Pseudocode of Step3 our VSE kernel}
\label{alg:vse_kernel}
\fontsize{7.5}{7.75}\selectfont

\begin{algorithmic}[1]
\Function{vse\_merge}{$in$, $out$, $left$, $mid$, $right$}
	\State{$l = left; r = mid; p = left; n_d = 0;$}
	\If{$mid - left < 16 || right - mid < 16$}
    	\While{$l < mid$ \&\& $r < right$}
    		\If{$input[r] < input[l]$} \Comment{correspond to variable $v$}
        		\State{$n_d += mid - l$; out[p++] = in[r++];}
       		\Else
        		\State{$out[p++] = in[l++]$;}
        	\EndIf
    	\EndWhile
	\Else
    
    	\Comment{count discordant pairs}
    	\While{$l < mid$ \&\& $r < right$}
    		\If{$in[r] < in[l]$}	\Comment{correspond to variable $v$}
        		\State{$n_d += mid - 1$; $r++$;}
        	\Else
        		\State{$l++$;}
        	\EndIf
		\EndWhile
    
    	\Comment{merge two sorted subarrays}
		\State{$l = left; r = mid;$}
    	\State{$vMin$ = \_mm512\_load\_epi32($in + l$); vMax = \_mm512\_load\_epi32($in + r$);}
        \State{$l += 16; r += 16;$}
    	\While{true}
    		\If{\_mm512\_reduce\_min\_epi32($vMin$) $\geq$ \_mm512\_reduce\_max\_epi32($vMax$)}
        		\State{\_mm512\_store\_epi32($out + p$, $vMax$); $p += 16; vMax = vMin;$}
       	 	\ElsIf{\_mm512\_reduce\_min\_epi32($vMax$) $\geq$ \_mm512\_reduce\_max\_epi32($vMin$)}
         		\State{\_mm512\_store\_epi32($out + p, vMin$); $p += 16; $}
       	 	\Else
            
            	\Comment{invoke Algorithm \ref{alg:bitonic_merge16}}
        		\State{{\tt bitonic\_merge\_16way}($vMin, vMax$);}
            	\State{\_mm512\_store\_epi32($out + p, vMin$); $p += 16;$}
        	\EndIf
        
			\If{$l + 16 \geq mid || r + 16 \geq right$}
    			\State{break};
    		\EndIf
    
    		\State{$A = in[l]; B = in[r];$ $C$ = \_mm512\_reduce\_max\_epi32($vMax$);}
   	 		\If{$C \leq A$ \&\& $C \leq B$}
      			\State{\_mm512\_store\_epi32($out + p, vMax$); $p += 16;$}
        		\State{$vMin$ = \_mm512\_load\_epi32($in + l$); $l += 16;$}
        		\State{$vMax$ = \_mm512\_load\_epi32($in + r$); $r += 16;$}
    		\ElsIf{$B < A$}
   				\State{$vMin$ = \_mm512\_load\_epi32($in + r$); $r += 16;$}
    		\Else
       			\State{$vMin$ = \_mm512\_load\_epi32($in + l$); $l += 16;$}
    		\EndIf
    	\EndWhile
    \EndIf
    
    \Comment{invoke Algorithm \ref{alg:bitonic_merge_leftover}}
    \State{{\tt bitonic\_merge\_16way\_leftover}($in, out, l, r, p, mid, right, vMax);$}

    \State \Return {$n_d$;}
\EndFunction

\vspace{1em}

\Function{kendall\_tau\_b\_step3\_vse}{$pairs$, $buffer$, $n$}
	\State{$n_d = 0; in = pairs; out = buffer;$}
    \For{$s$ = 1;  $s < n$; $s$ *= 2}
	    \For{$l$ = 0; $l <n$; $l$ += 2 * $s$}
        	\State{$m = \min(l + s, n)$;}
            \State{$r = \min(l + 2 * s, n)$;}
            
            \State{$n_d$ += {\tt vse\_merge}($in$, $out$, $l$, $m$, $r$);} \Comment{Perform out-of-place merge}
        \EndFor
        \State{{\tt swap}($in$, $out$);} \Comment{swap in and out}
  	\EndFor
    \If{pairs != in}
    	\State{{\tt memcpy}($pairs, in, n$ * sizeof(*$pairs$));}
    \EndIf
    
    \State \Return {$n_d$;}
\EndFunction
\end{algorithmic}
\end{algorithm}

In the literature, some research has been done to accelerate sorting algorithms by means of SIMD vectorized pairwise merge of sorted subarrays. Hiroshi \textit{et al.} \cite{inoue2007aa} employed a SSE vectorized odd-even merge network \cite{batcher1968sorting} to merge sorted subarrays in an out-of-core way. Chhugan \textit{et al.} \cite{chhugani2008efficient} adopted similar ideas to \cite{inoue2007aa}, but combined a SSE vectorized in-register odd-even merge sort \cite{batcher1968sorting} with an in-memory bitonic merge network. Chen \textit{et al.} \cite{xiaochen2013register} absorbed the merge-path idea of \cite{odeh2012merge, green2012gpu} and extended the SSE vectorized work of \cite{chhugani2008efficient} to take advantage of 512-bit SIMD VPUs on KNC Phis and used a vectorized in-register bitonic merge sort, instead of the in-register odd-even merge sort. In our VSE kernel, we engineered a 512-bit SIMD vectorized in-register bitonic merge network as the core of our pairwise merge procedure for sorted subarrays, which is similar to \cite{xiaochen2013register}, and further proposed a predict-and-skip mechanism to reduce the number of comparisons during pairwise merge.
\subsubsection{In-register bitonic merge network}
\label{sec:bitonimc_merge_16way}
\begin{figure}[!t]
\centering
\includegraphics[width=\linewidth]{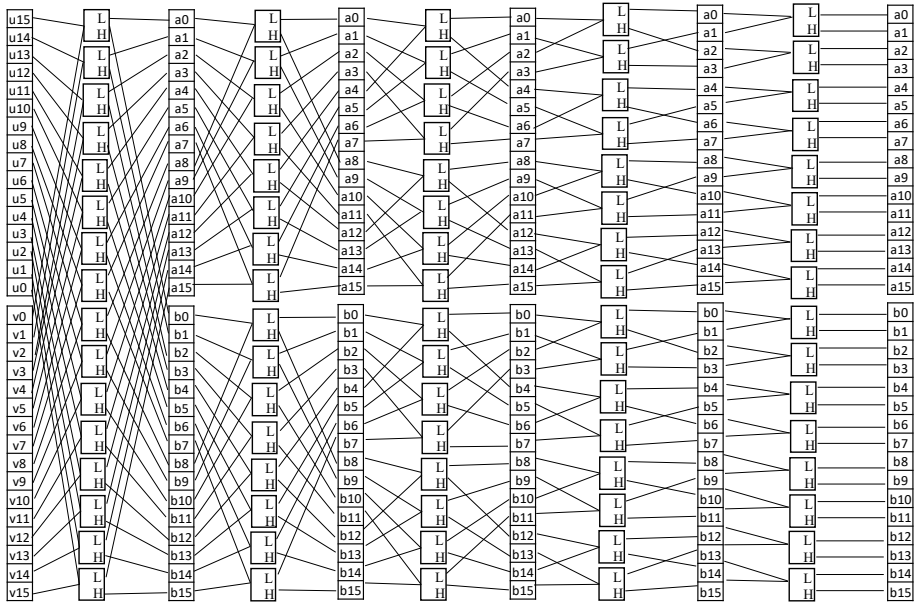}
\caption{16-way bitonic merge network: (1) input data are stored in two 16-lane vectors and (2) pipelining from left-to-right}
\label{fig:bitonic_merge16}
\end{figure}
The in-register bitonic merge network is the core of our vectorized pairwise merge of sorted subarrays. In our algorithm, this network has 16 ways and merges two sorted vectors $vMin$ and $vMax$ (Figure \ref{fig:bitonic_merge16} shows the computation layout of the network), where all elements in each vector are placed in ascending order from lane 0 to lane 15. In this case, to generate an input bitonic sequence from the two vectors, we need to reverse the order of all elements in one and only one vector (reverse $vMax$ in our case) and this order reversal is realized by one permutation instruction, i.e. \_mm512\_permutevar\_epi32($\cdot$). Having completed the order reversal, we can complete the sorting of vectors $vMin$ and $vMax$ in $\log_2(32)=5$ steps. Algorithm \ref{alg:bitonic_merge16} shows the pseudocode for our 16-way in-register bitonic merge network. From the code, the function {\tt bitonic\_merge\_16way} is composed of a fixed number of vector instructions and thus has a constant time complexity.
\begin{algorithm}
\centering
\caption{Pseudocode of our 16-way in-register bitonic merge network}
\label{alg:bitonic_merge16}
\fontsize{7.5}{7.75}\selectfont

\begin{algorithmic}[1]
\Procedure{bitonic\_merge\_16way}{$vMin$, $vMax$}

	\Comment{constant vector variables and reused}
	\State{//$vReverse$  = \_mm512\_set\_epi32(0, 1, 2, 3, 4, 5, 6, 7, 8, 9, 10, 11, 12, 13, 14, 15);}
    \State {//$vPermIndex16$ = \_mm512\_set\_epi32(7, 6, 5, 4, 3, 2, 1, 0, 15, 14, 13, 12, 11, 10, 9, 8)}
    \State {//$vPermIndex8$ = \_mm512\_set\_epi32(11, 10, 9, 8, 15, 14, 13, 12, 3, 2, 1, 0, 7, 6, 5, 4);}
    \State{//$vPermIndex4$ = \_mm512\_set\_epi32(13, 12, 15, 14, 9, 8, 11, 10, 5, 4, 7, 6, 1, 0, 3, 2);}
    \State{$//vPermIndex2$ = \_mm512\_set\_epi32(14, 15, 12, 13, 10, 11, 8, 9, 6, 7, 4, 5, 2, 3, 0, 1);}
    
    \Comment{Reserve vector $vMax$}
	\State{$vMax$ = \_mm512\_permutevar\_epi32($vReverse$, $vMax$);}
    
    \Comment{Level 1}
    \State{$vL1$ = \_mm512\_min\_epi32($vMin, vMax$);}
    \State{$vH1$ = \_mm512\_max\_epi32($vMin, vMax$);}
    
    \Comment {Level 2}
    \State{$vTmp$ = \_mm512\_permutevar\_epi32($vPermIndex16, vL1$);}
    \State{$vTmp2$ = \_mm512\_permutevar\_epi32($vPermIndex16, vH1$);}
    \State{$vL2$ = \_mm512\_mask\_min\_epi32($vL2$, 0x00ff, $vTmp, vL1$);}
	\State{$vH2$ = \_mm512\_mask\_min\_epi32($vH2$, 0x00ff, $vTmp2, vH1$);}
	\State{$vL2$ = \_mm512\_mask\_max\_epi32($vL2$, 0xff00, $vTmp, vL1$);}
    \State{$vH2$ = \_mm512\_mask\_max\_epi32($vH2$, 0xff00, $vTmp2, vH1$);}
    
	\Comment {Level 3}
    \State{$vTmp$ = \_mm512\_permutevar\_epi32($vPermIndex8, vL2$);}
    \State{$vTmp2$ = \_mm512\_permutevar\_epi32($vPermIndex8, vH2$);}
    \State{$vL3$ = \_mm512\_mask\_min\_epi32($vL3$, 0x0f0f, $vTmp, vL2$);}
	\State{$vH3$ = \_mm512\_mask\_min\_epi32($vH3$, 0x0f0f, $vTmp2, vH2$);}
	\State{$vL3$ = \_mm512\_mask\_max\_epi32($vL3$, 0xf0f0, $vTmp, vL2$);}
    \State{$vH3$ = \_mm512\_mask\_max\_epi32($vH3$, 0xf0f0, $vTmp2, vH2$);}
    
    \Comment {Level 4}
    \State{$vTmp$ = \_mm512\_permutevar\_epi32($vPermIndex4, vL3)$;}
    \State{$vTmp2$ = \_mm512\_permutevar\_epi32($vPermIndex4, vH3$);}
    \State{$vL4$ = \_mm512\_mask\_min\_epi32($vL4$, 0x3333, $vTmp, vL3$);}
	\State{$vH4$ = \_mm512\_mask\_min\_epi32($vH4$, 0x3333, $vTmp2, vH3$);}
	\State{$vL4$ = \_mm512\_mask\_max\_epi32($vL4$, 0xcccc, $vTmp, vL3$);}
    \State{$vH4$ = \_mm512\_mask\_max\_epi32($vH4$, 0xcccc, $vTmp2, vH3$);}
    
    \Comment {Level 5: vMin and vMax store the sorted sequence}
    \State{$vTmp$ = \_mm512\_permutevar\_epi32($vPermIndex2, vL4$);}
    \State{$vTmp2$ = \_mm512\_permutevar\_epi32($vPermIndex2, vH4$);}
    \State{$vMin$ = \_mm512\_mask\_min\_epi32($vMin$, 0x5555, $vTmp, vL4$);}
	\State{$vMax$ = \_mm512\_mask\_min\_epi32($vMax$, 0x5555, $vTmp2, vH4$);}
	\State{$vMin$ = \_mm512\_mask\_max\_epi32($vMin$, 0xaaaa, $vTmp, vL4$);}
    \State{$vMax$ = \_mm512\_mask\_max\_epi32($vMax$, 0xaaaa, $vTmp2, vH4$);}
\EndProcedure
\end{algorithmic}
\label{alg:bitonic_merge16}
\end{algorithm}
\begin{algorithm}
\centering
\caption{Pseudocode of processing the leftovers in both subarrays}
\label{alg:bitonic_merge_leftover}
\fontsize{7.5}{7.75}\selectfont

\begin{algorithmic}[1]
\Procedure{bitonic\_merge\_16way\_leftover}{$in, out, l, r, p, mid, right, vMax$}
	\State{$vIndexInc$ = \_mm512\_set\_epi32(15, 14, 13, 12, 11, 10, 9, 8, 7, 6, 5, 4, 3, 2, 1, 0);}
    \State{$vMaxInt$ = \_mm512\_set1\_epi32(0x7fffffff)} \Comment{maximum signed integer value}
    \State{$vMid$ = \_mm512\_set1\_epi32($mid$);}
    \State{$vRight$ = \_mm512\_set1\_epi32($right$);}
    
    \Comment{initialize the vector mask for $vMax$}
    \State{$maskMax$ = 0xffff;}
    \While{$l < mid || r < right$}
    	\If{$l < mid$ \&\& $r < right$}
        	\If{$in[r] < in[l]$}
                \State{$vTmp$ = \_mm512\_add\_epi32(\_mm512\_set1\_epi32($r$), $vIndexInc$)}
                \State{$maskMin$ = \_mm512\_cmplt\_epi32\_mask($vTmp$, $vRight$);}
                \State{$vMin$ = \_mm512\_mask\_load\_epi32($vMaxInt$, $maskMin$, $in + r$);  $r += 16$;}
          	\Else
                \State{$vTmp$ = \_mm512\_add\_epi32(\_mm512\_set1\_epi32($l$), $vIndexInc$)}
                \State{$maskMin$ = \_mm512\_cmplt\_epi32\_mask($vTmp$, $vMid$);}
                \State{$vMin$ = \_mm512\_mask\_load\_epi32($vMaxInt$, $maskMin$, $in + l$);  $l += 16$;}            	
            \EndIf
      	\ElsIf{$l < mid$}
        	\State{$vTmp$ = \_mm512\_add\_epi32(\_mm512\_set1\_epi32($l$), $vIndexInc$);}
        \State{$maskMin$ = \_mm512\_cmplt\_epi32\_mask($vTmp$, $vMid$);}
        \State{$vMin$ = \_mm512\_mask\_load\_epi32($vMaxInt$, $maskMin$, $in + l$);}
		\State{$tmp$ = \_mm512\_mask\_reduce\_max\_epi32($maskMax$, $vMax$); }
        \If{$tmp \leq$ \_mm512\_reduce\_min\_epi32($vMin$)}
          \State {break};
        \EndIf
        \State{$l += 16;$}
        \ElsIf{$r < right$}
         	\State{$vTmp$ = \_mm512\_add\_epi32(\_mm512\_set1\_epi32($r$), $vIndexInc$);}
        	\State{$maskMin$ = \_mm512\_cmplt\_epi32\_mask($vTmp$, $vRight$);}
        	\State{$vMin$ = \_mm512\_mask\_load\_epi32($vMaxInt, maskMin, in + r$);}
      	\State{$tmp$ = \_mm512\_mask\_reduce\_max\_epi32($maskMax, vMax$);}
        \If{$tmp \leq$ \_mm512\_reduce\_min\_epi32($vMin$)}
          \State{break};
        \EndIf
        \State{$r += 16;$}
        \EndIf
        
        \Comment{invoke Algorithm \ref{alg:bitonic_merge16}}
        \State{{\tt bitonic\_merge\_16way}($vMin, vMax$);}
        
        \Comment{calculate new vector masks}
        \State{$nb$ = \_mm\_countbits\_32($maskMin$) + \_mm\_countbits\_32($maskMax$);}
      \State{$maskMin = nb > 15 ?$ 0x0ffff : $(1 << nb) - 1;$}
      \State{$maskMax = nb < 17 ?$ 0 : $(1 << (nb - 16)) - 1;$}
      \State{\_mm512\_mask\_store\_epi32($out + p, maskMin, vMin$);}
      \State{$p$ += \_mm\_countbits\_32($maskMin$);}
    \EndWhile

	\Comment{write out $vMax$}
    \State{\_mm512\_mask\_packstorelo\_epi32($out + p, maskMax, vMax$);}
    \State{\_mm512\_mask\_packstorehi\_epi32($out + p + 16, maskMax, vMax$);}
    \State{$p$ += \_mm\_countbits\_32($maskMax$);}
    
    \Comment{copy out the rest}
	\If{$l < mid$}
		\State{{\tt memcpy}($out + p, in + l, (mid - l) *$ {\tt sizeof}(int));}
    \ElsIf{$r < right$}
    	\State{{\tt memcpy}($out + p, in + r, (right - r) *$ {\tt sizeof}(int));}
  	\EndIf
\EndProcedure
\end{algorithmic}
\end{algorithm}
\subsubsection{Vectorized pairwise merge of sorted subarrays}
Our vectorized pairwise merge of sorted subarrays relies on the aforementioned 16-way in-register bitonic merge network and adopted a very similar procedure to \cite{inoue2007aa}. Given two sorted subarrays $S_A$ and $S_B$, we assume that they are aligned to 64 bytes and their lengths $|S_A|$ and $|S_B|$ are multiples of 16, for the convenience of discussion. In this way, the vectorized pairwise merge works as follows.
\begin{enumerate}

\item loads the smallest 16 elements of $S_A$ and $S_B$ to $vMin$ and $vMax$, respectively, and advances the pointer of each subarray.

\item invokes {\tt bitonic\_merge\_16way($\cdot$)} to sort vectors $vMin$ and $vMax$, and then stores the content of $vMin$, the smallest 16 elements, to the resulting output array.

\item compares the next element of $S_A$ and $S_B$ and loads 16 elements into $vMin$ from the subarray whose next element is the smaller, followed by pointer advancing for the subarray.

\item jumps back to the second step and repeats the procedure until all elements in both subarrays have been processed.

\item completes the merge of $S_A$ and $S_B$ by writing the content of $vMax$ to the output array.
\end{enumerate}
Note that if $|S_A|$ or $|S_B|$ is not a multiple of 16, we will have to implement some special treatment to deal with the leftovers in $S_A$ and $S_B$. This special treatment is implemented as Algorithm \ref{alg:bitonic_merge_leftover} and invoked by Algorithm \ref{alg:vse_kernel} (see line 46). From the above procedure, we can see that the time complexity of merging $S_A$ and $S_B$ is linear and thus the VSE kernel has a time complexity of $O(n\log n)$.

As assumed that both $S_A$ and $S_B$ are aligned to 64 bytes, we can therefore use only one {\tt \_mm512\_load\_epi32($\cdot$)} instruction to load 16 integer elements from memory to a vector and only one {\tt \_mm512\_store\_epi32($\cdot$)} instruction to store 16 elements in a vector to memory. This is because memory address is guaranteed to keep aligned to 64 bytes during pairwise merge. Defining $T_{rd}$ to denote the latency of memory load (in clock cycles) and $T_{wt}$ to denote the latency of memory store, the number of clock cycles per element load from memory can be estimated as $(T_{rd}+16)/16=T_{rd}/16 + 1$ clock cycles and the number of clocks cycles per element store to memory as $(T_{wt}+16)/16=T_{wt}/16 + 1$ clock cycles. Moreover, {\tt bitonic\_merge\_16way($\cdot$)} is composed of 27 vector instructions (see lines 7$\sim$33 in Algorithm \ref{alg:bitonic_merge16}) and each of its invocations leads to the output of 16 elements. In these regards, the average number of instructions per element can be estimated as $27/16\approx1.69$. Considering that we only used simple integer and single-source permutation instructions, each of which has a two clock cycle latency according to the Intel Xeon Phi processor software optimization manual \cite{phioptimization2016}, the average computational cost per element can be estimated as $\lceil 2\times 27/16\rceil = 4$ clock cycles in the worst cases where consecutive instructions always have data dependency. In order to reduce inter-instruction latency caused by data dependency, we have manually tuned the order of instructions by means of interleaving. Since each instruction has a latency of two clock cycles, this interleaving could enable to execute one instruction in only one clock cycle, thus further reducing the cost per element to $\lceil 1\times 27/16\rceil = 2$ clock cycles optimistically. In sum, the overall cost per element can be favorably estimated to be $(T_{rd} + T_{wr})/16 + 4$ clock cycles.

We have proposed a predict-and-skip scheme to determine ($i$) whether we need to invoke the 16-way bitonic merge network and ($ii$) whether we should load two vectors of new elements from $S_A$ and $S_B$, respectively. For cases ($i$), the rationale is if the minimum value in $vMin$ ($vMax$) is $\geq$ the maximum value in $vMax$ ($vMin$), every value in $vMin$ ($vMax$) will be $\geq$ to all values in $vMin$ ($vMax$). In this case, there is no need of invoking the bitonic merge network (lines 23$\sim$26 in Algorithm \ref{alg:vse_kernel}), since the order of all elements in both vectors is already deterministic. For cases ($ii$), we compare the maximum value $C$ in $vMax$ with both the smallest element $A$ in the rest of $S_A$ and the one $B$ in the rest of $S_B$ (lines 34$\sim$38 in Algorithm \ref{alg:vse_kernel}). If $C\leq A$ and $C\leq B$, none of the elements in $vMax$ will be greater than any element in the rest of $S_A$ and $S_B$, indicating that the absolute order of all $vMax$ elements has already been determined. In this case, we write $vMax$ to the resulting output array and load two vectors of new elements from $S_A$ and $S_B$ to $vMin$ and $vMax$, respectively. The updates in $vMin$ and $vMax$ will be used for the next iteration of comparison.

It is worth mentioning that we also developed a 32-way bitonic merge network to implement the VSE kernel. Unfortunately, this new kernel based on 32-way merge network yielded slightly inferior performance to the kernel with the 16-way network through our tests. In this regard, we adopted the 16-way merge network in our VSE kernel.
\section{Parallel All-Pairs Computation}
\subsection{All-Pairs Computation Framework}
We consider the $m\times m$ job matrix to be a 2-dimensional coordinate space on the Cartesian plane, and define the left-top corner to be the origin, the horizontal $x$-axis (corresponding to columns) in left-to-right direction and the vertical $y$-axis (corresponding to rows) in top-to-bottom direction. For non-symmetric all-pairs computation (non-commutative pairwise computation), the workload distribution over processing elements (PEs), e.g. threads, processes, cores and etc., would be relatively straightforward. This is because coordinates in the 2-dimensional matrix corresponds to distinct jobs. Unlike non-symmetric all-pairs computation, it suffices by only computing the upper-triangle (or lower-triangle) of the job matrix for symmetric all-pairs computation (commutative pairwise computation). In this case, balanced workload distribution could be more complex than non-symmetric all-pairs computation.

\begin{figure}
\centering
\begin{subfigure}[b]{0.23\linewidth}
\centering
\includegraphics[width=\linewidth]{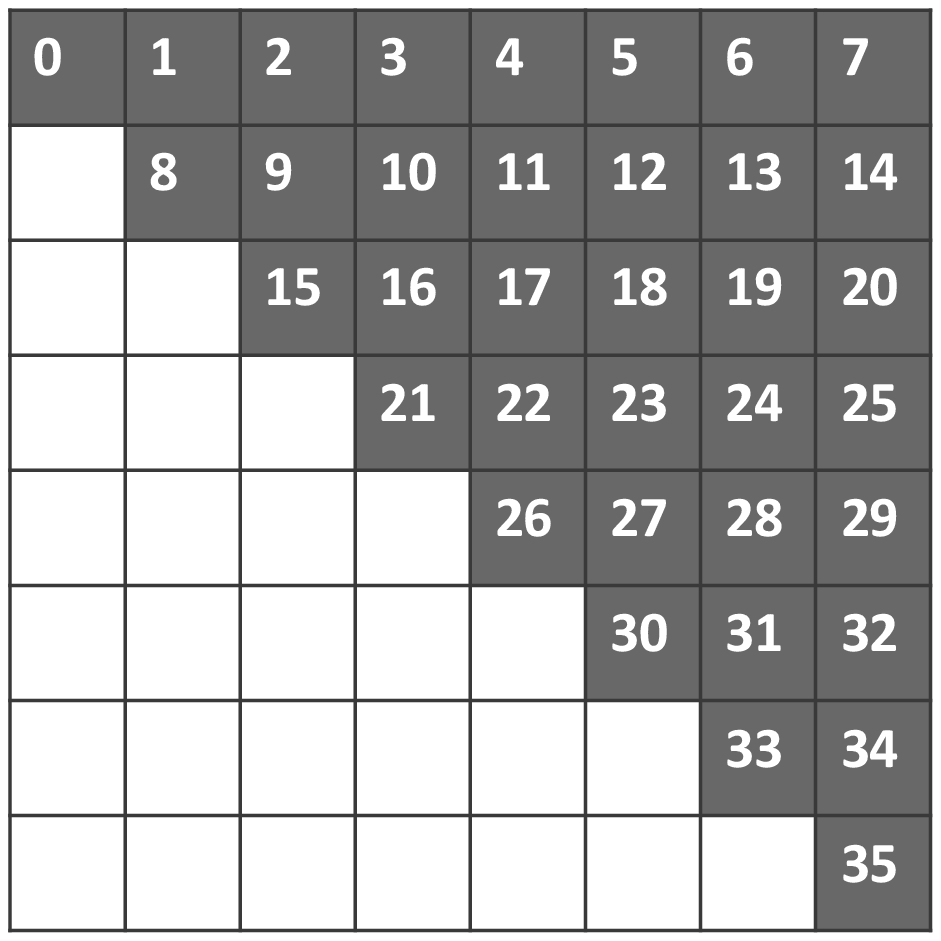}
\caption{}
\label{fig:direct}
\end{subfigure}
\begin{subfigure}[b]{0.34\linewidth}
\centering
\includegraphics[width=\linewidth]{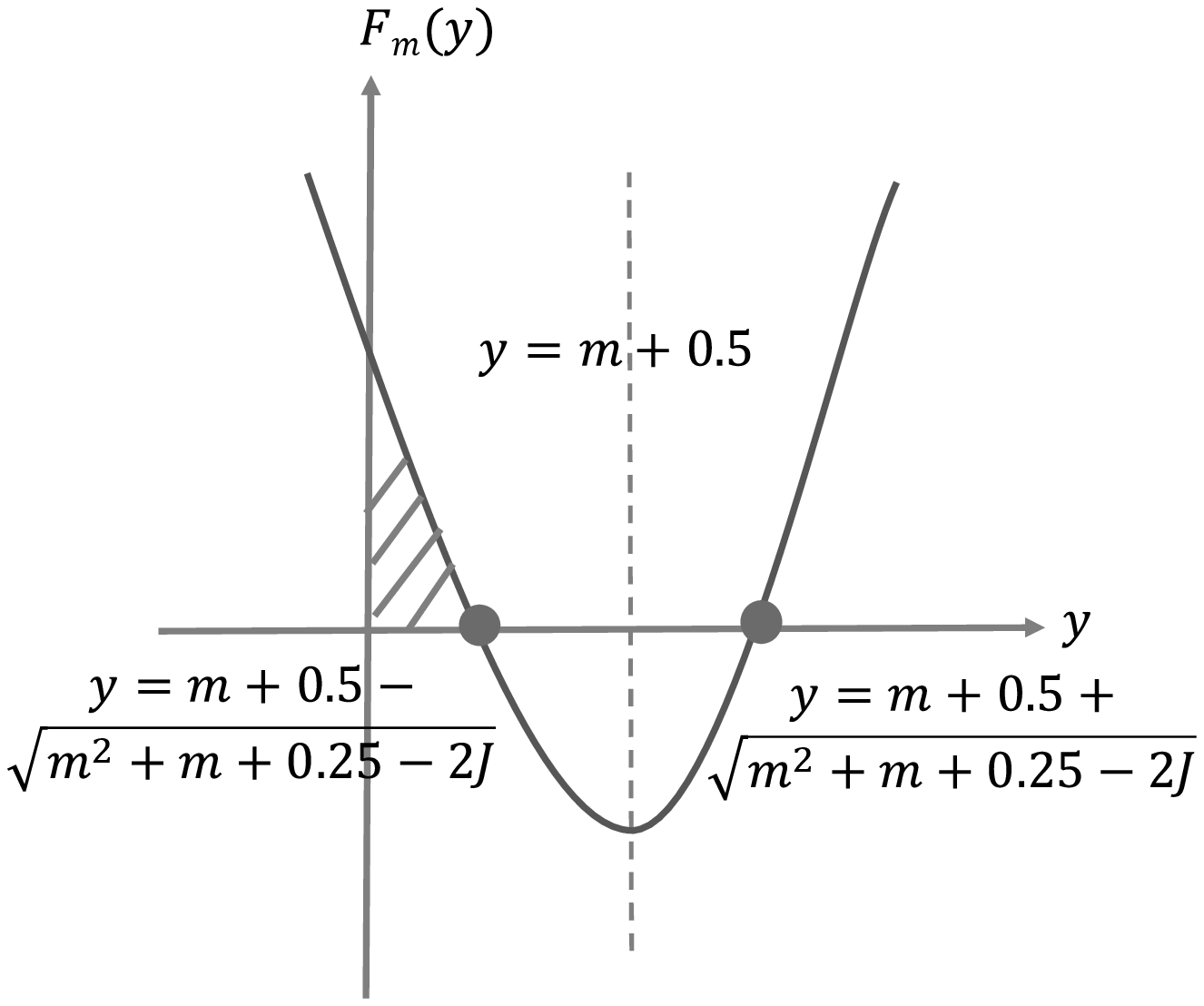}
\caption{}
\label{fig:left_region}
\end{subfigure}
\begin{subfigure}[b]{0.41\linewidth}
\centering
\includegraphics[width=\linewidth]{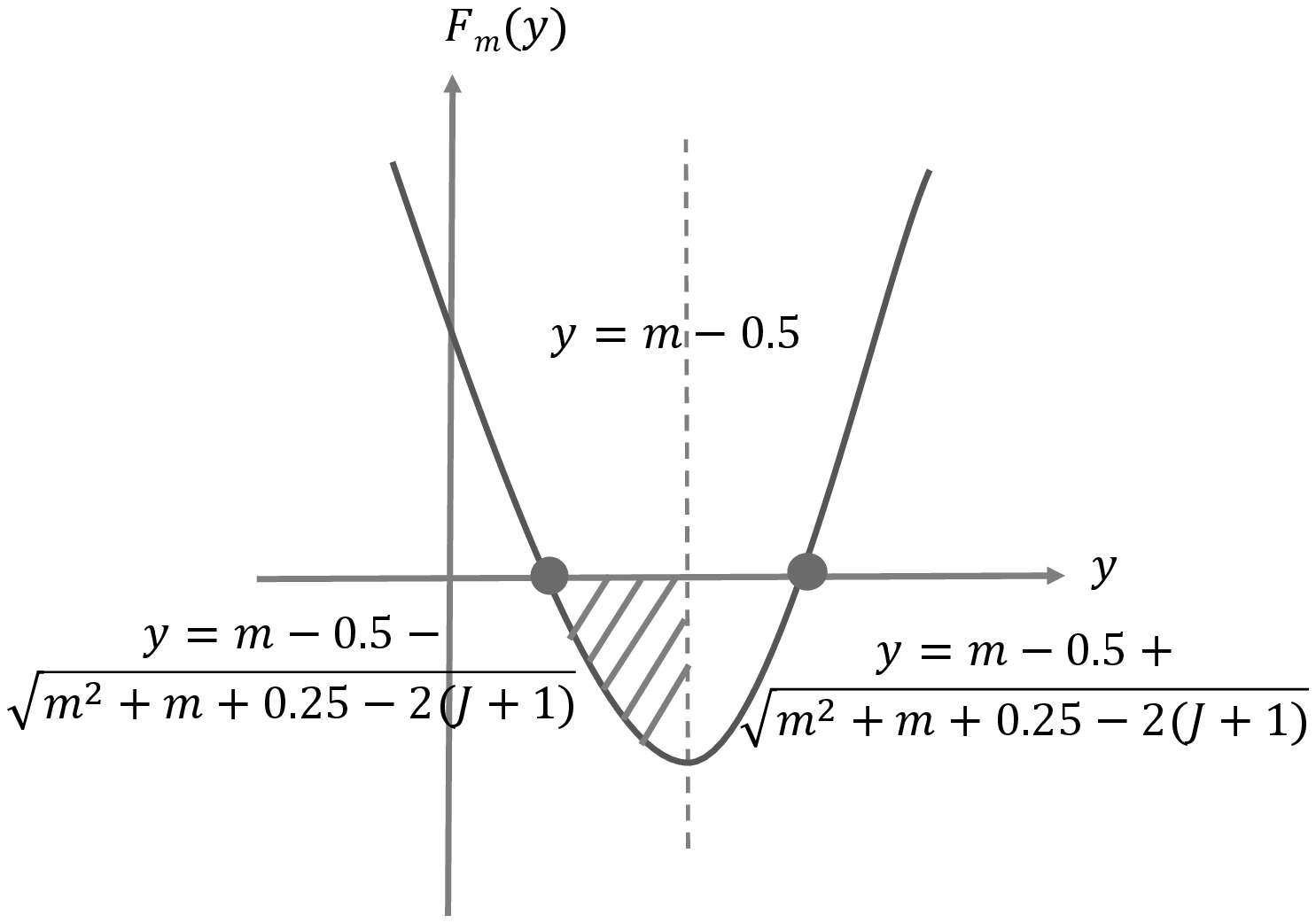}
\caption{}
\label{fig:right_region}
\end{subfigure}
\caption{(a) an example direct bijective mapping between job identifier and coordinate space, (b) solution region for Equation (\ref{equation:inequality_y_hi}), and (c) solution region for Equation (\ref{equation:inequality_y_low})}
\end{figure}

In this paper, we propose a generic framework for workload balancing in symmetric all-pairs computation, since the computation of the $\tau$ coefficient between any $u$ and $v$ is commutable as mentioned above. This framework works by assigning each job a unique identifier and then building a bijective relationship between a job identifier $J_m(y, x)$ and its corresponding coordinate $(y, x)$. We refer to this as {\tt direct bijective mapping}. While facilitating balanced workload distribution, this mapping merely relies on bijective functions, which is a prominent feature distinguished from existing methods. To the best of our knowledge, in the literature bijective functions have not ever been proposed for workload balancing in symmetric all-pairs computation. In \cite{kiefer2010pairwise}, the authors used a very similar job numbering approach to ours (shown in this study), but did not derive a bijective function for symmetric all-pairs computation. Our framework can be applied to cases with identical (e.g. using static workload distribution) or varied workload per job (e.g. using a shared integer counter to realize dynamic workload distribution via remote memory access operations in MPI \cite{gropp2014using} and Unified Parallel C (UPC) programming models \cite{el2006upc} \cite{gonzalez2016parallel}) and is also particularly useful for parallel computing architectures with hardware schedulers such as GPUs and FPGAs. In the following, without loss of generality, we will interpret our framework relative to the upper triangle of the job matrix by counting in the major diagonal. Nonetheless, this framework can be easily adapted to the cases excluding the major diagonal.
\paragraph{Direct bijective mapping}
Given a job $(y, x$) in the upper triangle, we compute its integer job identifier $J_m(y, x)$ as
\begin{equation}
\begin{array}{ll}
J_m(y, x) = F_m(y) + x - y,&0\leq y\leq x < m\\
\end{array}
\label{equation:j}
\end{equation}
for dimension size $m$. In this equation, $F_m(y)$ is the total number of cells preceding row $y$ in the upper triangle and is computed as
\begin{equation}
F_m(y) = \frac{y(2m-y+1)}{2}
\label{equation:f}
\end{equation}
where $y$ varies in $[0, m]$. In this way, we have defined Equation (\ref{equation:j}) based on our job numbering policy, i.e. all job identifiers vary in \mbox{$[0, m(m+1)/2)$} and jobs are sequentially numbered left-to-right and top-to-bottom in the upper triangle (see Fig. \ref{fig:direct} for an example).

Reversely, given a job identifier $J = J_m(y, x)$ ($0\leq J < m(m+1)/2$), we need to compute the coordinate $(y, x)$ in order to locate the corresponding variable pair. As per our definition, we have
\begin{equation}
\left\{
\begin{array}{l}
J \geq F_m(y) \Leftrightarrow y^2 - (2m+1)y + 2J \geq 0\\
J \leq F_m(y+1) - 1 \Leftrightarrow y^2 - (2m-1)y + 2(J+1) - 2m \leq 0\\
\end{array}
\right.
\label{equation:inequality_j}
\end{equation}
By solving $J\geq F_m(y)$ (see Figure \ref{fig:left_region}), we get
\begin{equation}
y \leq m + 0.5 - \sqrt{m^2 + m + 0.25 - 2J}
\label{equation:inequality_y_hi}
\end{equation}
while getting
\begin{equation}
y \geq m - 0.5 - \sqrt{m^2 + m + 0.25 - 2(J+1)}
\label{equation:inequality_y_low}
\end{equation}
by solving $J \leq F_m(y+1) - 1$ (see Figure \ref{fig:right_region}). In this case, by defining $\Delta = \sqrt{m^2 + m + 0.25 - 2(J+1)}$, $\Delta' = \sqrt{m^2 + m + 0.25 - 2J}$ and $z = m - 0.5 - \sqrt{m^2 + m + 0.25 - 2(J+1)}$, we can reformulate Equations (\ref{equation:inequality_y_hi}) and (\ref{equation:inequality_y_low}) to be $z \leq y \leq z + 1 + \Delta - \Delta'$. Because $\Delta < \Delta'$, we know that $[z, z + 1 + \Delta - \Delta']$ is a sub-range of $[z, z+1)$ and thereby have $z \leq y < z + 1$. Based on our job numbering policy stated before, as a function of integer $y$, Equation (\ref{equation:inequality_j}) definitely has $y$ solutions, meaning that at least one integer exists in 
$[z, z + 1 + \Delta - \Delta']$, which satisfies Equation (\ref{equation:inequality_j}). Meanwhile, it is known that there always exists one and only one integer in $[z, z+1)$ (can be easily proved) and this integer equals $\lceil z\rceil$, regardless of the value of $z$. Since $[z, z + 1 + \Delta - \Delta']$ is a sub-range of $[z, z + 1)$, we can conclude that Equation (\ref{equation:inequality_j}) has a unique solution $y$ that is computed as
\begin{equation}
	y = \lceil z \rceil = \Bigl\lceil m - 0.5 - \sqrt{m^2 + m + 0.25 - 2(J+1)} \Bigr\rceil
\label{equation:y}
\end{equation}
Having got $y$, we can compute the coordinate $x$ as
\begin{equation}
x = J + y - F_m(y)
\label{equation:x}
\end{equation}
based on Equation (\ref{equation:j}).
\subsection{Tiled Computing}
Based on the direct bijective mapping, we adopted tiled computing with the intention to benefit from L1/L2 caches. The rationale behind the tiled computing is loading into cache a small subset of the bigger dataset and reusing this block of data in cache for multiple passes. This technique partitions a matrix into a non-overlapping set of equal-sized $q\times q$ tiles. In our case, we partition the job matrix and produce a tile matrix of size $w\times w$ tiles, where $w=\lceil m/q\rceil$. In this way, all jobs in the upper triangle of the job matrix are still fully covered by the upper triangle of the tile matrix. By treating a tile as a unit, we can assign a unique identifier to each tile in the upper triangle of the tile matrix and then build bijective functions between tile identifiers and tile coordinates in the tile matrix, similarly as we do for the job matrix.

Because the tile matrix has an identical structure to the original job matrix, we can directly apply our bijective mapping to the tile matrix. In this case, given a coordinate $(y_q, x_q)$ ($0\leq y_q \leq x_q < w$) in the upper triangle of the tile matrix, we can compute a unique tile identifier $J_w(y_q, x_q)$ as
\begin{equation}
\begin{array}{ll}
J_w(y_q, x_q) = F_w(y_q) + x_q - y_q,& 0\leq y_q \leq x_q < w\\
\end{array}
\label{equation:jt}
\end{equation}
where $F_w(y_q)$ is defined similar to Equation (\ref{equation:f}) as
\begin{equation}
F_w(y_q) = \frac{y_q(2w - y_q + 1)}{2}
\label{equation:ft}
\end{equation}
Likewise, given a tile identifier $J_w$, such that $0\leq J_w < w(w+1)/2$, we can reversely compute its unique vertical coordinate $y_q$ as
\begin{equation}
y_q =  \Bigl\lceil w - 0.5 - \sqrt{w^2 + w + 0.25 - 2(J_w+1)}\Bigl\rceil
\label{equation:yt}
\end{equation}
and subsequently its unique horizontal coordinate $x_q$ as
\begin{equation}
x_q = J_w + y_q - F_w(y_q)
\label{equation:xt}
\end{equation}

As title size is subject to cache sizes and input data, tuning tile size is believed to be important for gaining high performance. This tuning process, however, is tedious and has to be conducted case-by-case. For convenience, we empirically set $q$ to 8 for CPUs and to 4 for Phis. Nevertheless, users can feel free to tune tile sizes to meet their needs.
\subsection{Multithreading}
\subsubsection{Asynchronous kernel execution}
When $m$ is large, we may not have sufficient memory to reside the resulting $m\times m$ correlation matrix $M_\tau$ entirely in memory. To overcome memory limitation, we adopted a multi-pass kernel execution model which partitions the tile identifier range $[0, w(w+1)/2)$ into a set of non-overlapping sub-ranges (equal-sized in our case) and finishes the computation one sub-range after another. In this way, we do not need to allocate the whole memory for matrix $M_\tau$. Instead, we only need to allocate a small amount of memory to store the computing results of one sub-range, thus considerably reducing memory footprint.

When using the KNC Phi, we need to transfer the newly computed results from the Phi to the host after having completed each pass of kernel execution. If kernel execution and data transfer is conducted in sequential, the Phi will be kept idle during the interim of device-to-host data transfer. In this regard, a more preferable solution would be to employ asynchronous kernel execution by enabling concurrent execution of host-side tasks and accelerator-side kernel execution. Fortunately, KNC Phis enable such a kind of asynchronous data transfer and kernel execution associated with the offload model. This asynchronous execution can be realized by coupling the {\tt signal} and {\tt wait} clauses, both of which are associated with each other via a unique identifier. That is, a {\tt signal} clause initiates an asynchronous operation such as data transfer and kernel execution, while a {\tt wait} clause blocks the current execution until the associated asynchronous operation is completed. Refer to our previous work \cite{liu2016parallel} for more details about the host-side asynchronous execution workflow proposed.
\subsubsection{Workload balancing}
For the variant using the na\"{i}ve kernel, all jobs have the same amount of computation (refer to Algorithm \ref{alg:naive_kernel}). Thus, given a fixed number of threads, we evenly distribute jobs over the set of threads by assigning a thread to process one tile at a time. In contrast, for the two variants using the sorting-enabled kernels, different jobs may have different amount of computation because of the two rounds of sort used in Step1 and Step3. Typically, an OpenMP dynamic scheduling policy is supposed to be in favor to address workload irregularity, albeit having relatively heavier workload distribution overhead than static scheduling. However, it is observed that dynamic scheduling produces slightly inferior performance to static scheduling through our tests. In this regard, we adopted static scheduling in our two sorting-enabled variants and assigned one thread to process one tile at a time, same as the na\"{i}ve variant.
\subsection{Distributed Computing}
On KNC Phi clusters, we used MPI offload model which launches MPI processes just as an ordinary CPU cluster does. In this model, one or more Phis will be associated to a parental MPI process, which utilizes offload pragmas/directives to interact with the affiliated Phis, and inter-Phi communications need to be explicitly managed by parental processes. In this sense, Phis may not perceive the existence of remote inter-process communications. Our distributed implementations require one-to-one correspondence between MPI processes and Phis, and adopted a static workload distribution scheme based on tiled computation. This static distribution is inspired by our practice in multithreading on single Phis. In our implementation, given $p$ processes, we evenly distribute tiles onto the $p$ processes with the $i$-th ($0\leq i < p$) process assigned to compute the tiles whose identifiers are in $[i\times \lceil \frac{w(w+1)}{2p}\rceil, (i + 1)\times \lceil \frac{w(w+1)}{2p}\rceil)$. Within each process, we adopted asynchronous execution workflow as well.
\section{Performance Evaluation}
We assessed the performance of our algorithm from four aspects: ($i$) comparison between our three variants, ($ii$) comparison with widely used counterparts: MATLAB (version R2015b) and R (version 3.2.0), ($iii$) multithreading scalability on a single Phi, and ($iv$) distributed computing scalability on Phi clusters. In these tests, we used four real whole human genome gene expression datasets (refer to Table \ref{tab:dataset}) produced by Affymetrix Human Genome U133 Plus 2.0 Array. These datasets are publicly available in the GPL570 data collection of SEEK \cite{zhu2015targeted}, a query-based computational gene co-expression search engine over large transcriptomic databases. In this study, unless otherwise  specified, we compute $\tau$ coefficients between genes, meaning that $m$ is equal to the number of genes and $n$ equal to the number of samples for each dataset.

All tests are conducted on 8 compute nodes in CyEnce HPC Cluster (Iowa State University), where each node has two high-end Intel E5-2650 8-core 2.0 GHz CPUs, two 5110P Phis (each has 60 cores and 8 GB memory) and 128 GB memory. Our algorithm is compiled by Intel C++ compiler v15.0.1 with option {\tt -fast} enabled. In addition, for distributed computing scalability assessment, when two processes are launched into the same node, we used the environment variable {\tt I\_MPI\_PIN\_PROCESSOR\_LIST} to guide Intel MPI runtime system to pin the two processes within the node to distinct CPUs (recall that one node has two CPUs).
\begin{table}
\centering
\caption{Information of whole human genome gene expression datasets used}
\label{tab:dataset}
\begin{tabular}{llll}
\hline
Name&	No. of Genes&	No. of Samples&	Platform\\
\hline
DS21050&		17,941&		310&	Affymetrix\\
DS19784&		17,941&		320&	Affymetrix\\
DS13070&		17,941&		324&	Affymetrix\\
DS3526&		17,941&		353&	Affymetrix\\
\hline
\end{tabular}
\end{table}
\subsection{Assessment of Our Three Variants}
All of the three variants work on CPUs, Phis and their clusters. For the na\"{i}ve and GSE variants, they both used the same C++ core code for CPU- and MIC-oriented instances. For the VSE variant, its 512-bit SIMD vectorization is only applicable to Phis. In this regard, to support CPUs, we further developed a non-vectorized merge sort in Step1 and a non-vectorized discordant pair counting algorithm in Step3 instead, based on the aforementioned packed integer representation. Considering that this non-vectorized version also works on Phis, we have examined how much our 512-bit SIMD vectorization contributes to speed improvement by comparing the vectorized version with the non-vectorized one on the same 5110P Phi. Interestingly, by measuring the correlation between genes using the datasets in Table \ref{tab:dataset}, we observed that the non-vectorized version is on a par with the vectorized one. This may be due to the relatively small value of $n$, which is only 353 at maximum. Based on this consideration, we further evaluated both versions by measuring the correlation between samples, where $m$ will be equal  to the number of samples and $n$ equal to the number of genes. In this case, $n$ becomes 17,941 for each dataset, more than $50\times$ larger than before. In this context, our performance assessment exposed that the vectorized version is consistently superior to the non-vectorized one, yielding very stable speedups averaged to be 1.87 for all datasets. This result was exciting, proving that our 512-bit SIMD vectorization did boost speed even for moderately large $n$ values. Hence, we have used the vectorized version of the VSE variant for performance measurement all throughout our study.
\subsubsection{On CPU} 
\begin{figure}
\centering
\includegraphics[width=0.8\linewidth]{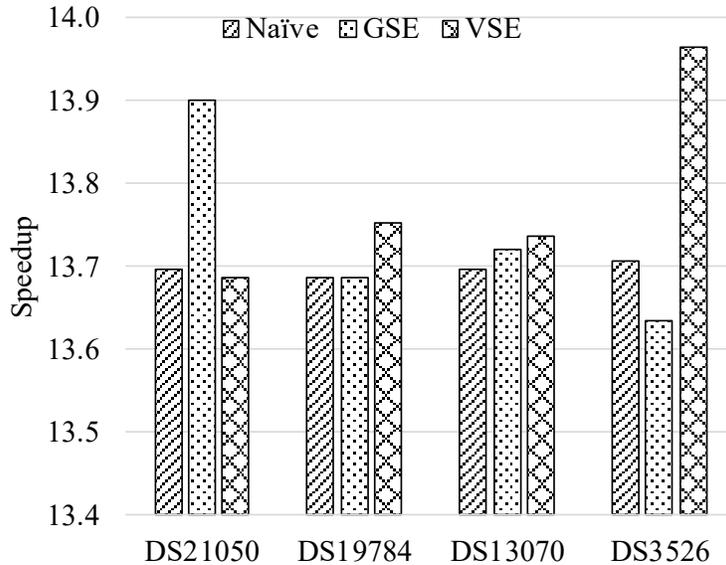}
\caption{Speedups of 16-threaded instances over single-thread ones on multiple CPU cores}
\label{fig:cpu_speedup16t}
\end{figure}

We first compared the performance of our three variants on multiple CPU cores. Table \ref{tab:cpu_performance} shows the performance of each variant on 16 CPU cores and Figure \ref{fig:cpu_speedup16t} shows the speedup of the 16-threaded instance of each variant over its single-threaded one.

For the na\"{i}ve variant, its 16-threaded instance achieves a roughly constant speedup of 13.70 over its single-threaded one. This observation is consistent with our expectation, as the runtime of the na\"{i}ve kernel is subject to the vector size $n$ but independent of actual vector content (refer to the implementation shown in Algorithm \ref{alg:naive_kernel}). In contrast, the speed of the two sorting-enabled kernels are sensitive to vector content to some degree. This can be explained by the following three factors: variable pair sorting in Step1, discordant pair counting based on pairwise merge of sorted subarrays in Step3, and linear-time scans for determining tie groups in Step2 and Step4. Nevertheless, for both sorting-enabled variants, their 16-threaded instances yield relatively consistent speedups over their single-threaded ones, respectively. For the GSE (VSE) variant, its 16-threaded instance runs $13.74\times$ ($13.79\times$) faster than its single-threaded one on average, with a minimum speedup 13.63 (13.69) and a maximum speedup 13.90 (13.97). For each case, the two sorting-enabled variants both yield superior performance to the na\"{i}ve variant, where the average speedup is 1.60 for the GSE kernel and 2.72 for the VSE kernel.
\begin{table}
\centering
\caption{Performance comparison of three variants on multiple CPU cores}
\label{tab:cpu_performance}
\begin{tabular}{lllllll}
\hline
\multirow{2}{*}{Instance}&	\multirow{2}{*}{Dataset}&	\multicolumn{3}{c}{Time (s)}&	\multicolumn{2}{c}{Speedup}\\
\hhline{~~-----}
&&	Na\"{i}ve&	GSE&	VSE&	GSE&	VSE\\
\hline
\multirow{4}{*}{Single-threaded}
	&DS21050&	10,052 & 6,609 & 3,758&	1.52&	2.67\\
	&DS19784&	10,690 & 6,776 & 3,938&		1.58&	2.71\\
	&DS13070&	10,975 & 6,920 & 4,041&		1.59&	2.72\\
	&DS3526&	12,993 & 7,619 & 4,742&	 	1.71&	2.74\\
\hline
\multirow{4}{*}{16-threaded}
	&DS21050&	734 & 475 & 275&	1.54&	2.67\\
	&DS19784&	781 & 495 & 286	&	1.58&	2.73\\
	&DS13070&	801 & 504 & 294	&	1.59&	2.72\\
	&DS3526&	948 & 559 & 340	&	1.70&	2.79\\
\hline
\end{tabular}
\end{table}
\subsubsection{On Xeon Phi}
Subsequently, we evaluated the three variants on a single 5110P Phi (see Table \ref{tab:phi_performance}). From Table \ref{tab:phi_performance}, the VSE variant outperforms the na\"{i}ve variant for each dataset by yielding an average speedup of 1.60 with the minimum speedup 1.53 and the maximum speedup 1.71. The GSE variant is superior to the na\"{i}ve one on the DS3526 dataset, but inferior to the latter on the remaining three datasets. Interestingly, by revisiting the CPU results shown in Table \ref{tab:cpu_performance}, we recalled that the former actually performs consistently better than the latter for each dataset on CPU. Since both the na\"{i}ve and the GSE variants use the same pairwise $\tau$ coefficient kernel code for their corresponding CPU- and MIC-oriented implementations, this discordant performance ranking between the two variants could owe to the architectural differences of the two types of processing unit (PU). For instance, by enabling auto-vectorization, the na\"{i}ve kernel (see Algorithm \ref{alg:naive_kernel}) can concurrently process 16 integer elements by one 512-bit SIMD vector instruction, in contrast with 4 integer elements by one 128-bit SSE instruction on CPU. This fourfold increase in parallelism would enable the na\"{i}ve kernel to further boost performance.
\begin{table}
\centering
\caption{Performance comparison of the three variants on Phi}
\label{tab:phi_performance}
\begin{tabular}{llllll}
\hline
\multirow{2}{*}{Dataset}&	\multicolumn{3}{c}{Time (s)}&	\multicolumn{2}{c}{Speedup}\\
\hhline{~-----}
&Na\"{i}ve&	GSE&	VSE&	GSE&	VSE\\
\hline
DS21050&	400& 	419& 	261&	0.95&	1.53\\
DS19784&	424& 	434& 	266&	0.98&	1.59\\
DS13070&	433& 	459& 	276&	0.94&	1.57\\
DS3526&		506& 	488& 	296&	1.04&	1.71\\
\hline
\end{tabular}
\end{table}

\begin{figure}
\centering
\begin{subfigure}[b]{0.495\linewidth}
\centering
\includegraphics[width=\linewidth]{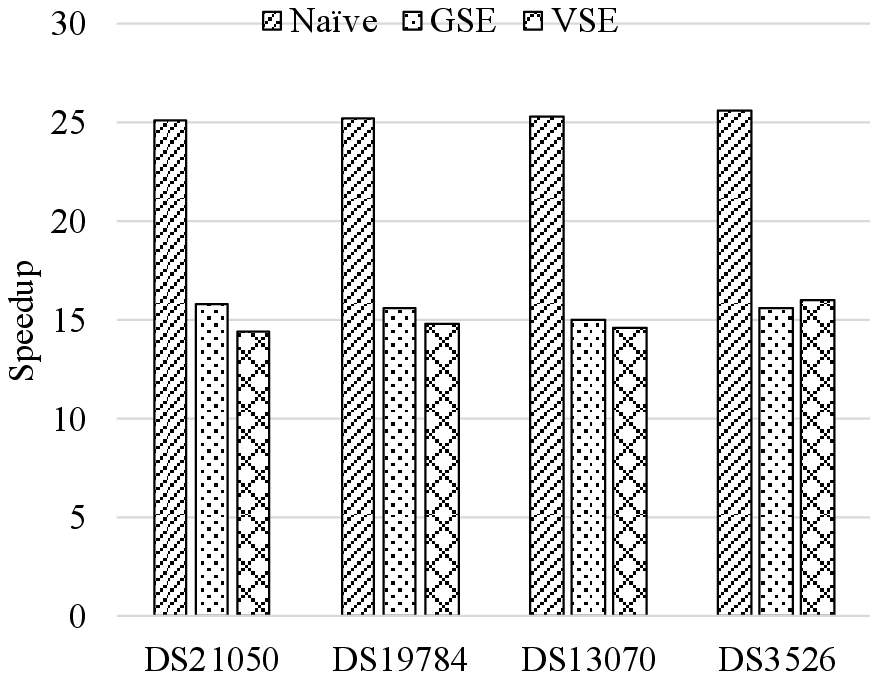}
\caption{}
\label{fig:phi_over_1t}
\end{subfigure}
\begin{subfigure}[b]{0.495\linewidth}
\centering
\includegraphics[width=\linewidth]{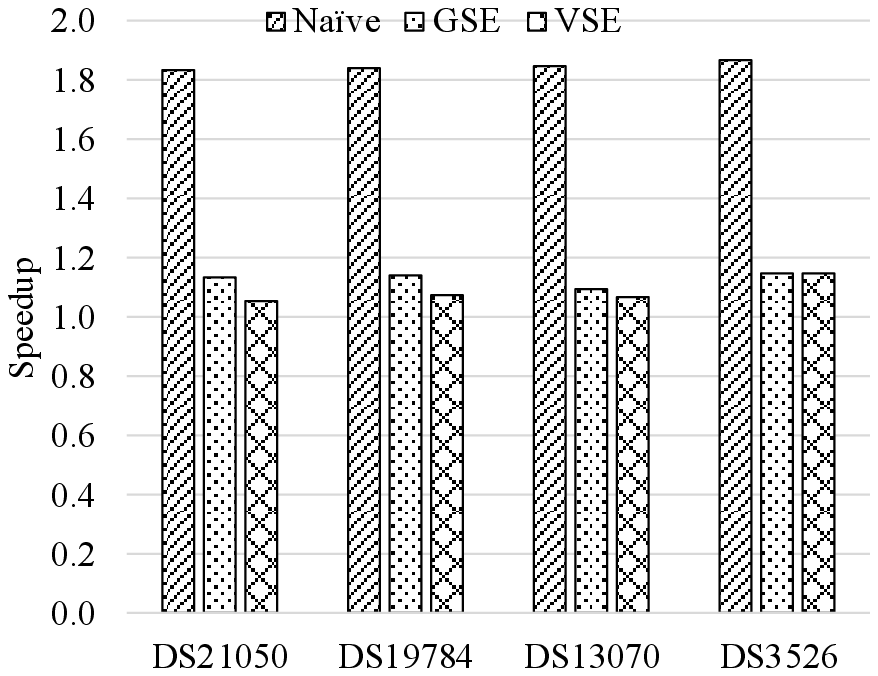}
\caption{}
\label{fig:phi_over_16t}
\end{subfigure}
\caption{Speedups of each variant on Phi over on CPU: (a) speedups over one CPU core and (b) speedups over 16 CPU cores.}
\label{fig:phi_over_cpu}
\end{figure}

Figure \ref{fig:phi_over_cpu} shows the speedups of each variant on the Phi over on multiple CPU cores for each dataset. From Figure \ref{fig:phi_over_cpu}, it is observed that the Phi instance of each variant outperforms its corresponding single-threaded and 16-threaded CPU instances for each dataset. Specifically, the Phi instance of the na\"{i}ve variant runs $25.34\times$ faster on average than its single-threaded CPU instance, with the maximum speedup 25.67, and $1.85\times$ faster on average than the 16-threaded CPU instance, with the maximum speedup 1.87. Compared to their single-threaded CPU instances, the Phi instances of the GSE and VSE variants produce the average speedups of 15.52 and 14.96 and the maximum speedups of 15.77 and 16.00, respectively. Meanwhile, in comparison to their 16-threaded CPU instances, their Phi instances performs $1.13\times$ and $1.08\times$ better on average and $1.15\times$ and $1.15\times$ better at maximum, respectively.
\subsection{Comparison With MATLAB and R}
\label{sec:matlab_r}
Secondly, we compared our algorithm with all-pairs $\tau$ coefficient implementations in the widely used MATLAB (version R2015b) and R (version 3.2.0). MATLAB and R are executed in one 16-core compute node mentioned above, where MATLAB runs 16 threads as it supports multithreading and R runs in sequential. For fair comparison, both MATLAB and R merely compute all-pairs $\tau$ coefficient values, without conducting statistical tests, same as our algorithm does. Table \ref{tab:matlab_r} shows the runtimes of the 16-threaded MATLAB and sequential R instances as well as their comparison with our three variants that execute on the 5110P Phi. From the table, each variant demonstrates excellent speedups, i.e. two orders-of-magnitude over 16-threaded MATLAB and three orders-of-magnitude over sequential R. Compared to 16-threaded MATLAB, the average and maximum speedups are 472 and 484 for the na\"{i}ve variant, 462 and 494 for the GSE variant and 756 and 812 for the VSE variant, respectively. In contrast, the average and maximum speedups over sequential R are 671 and 683 for the na\"{i}ve variant, 656 and 709 for the GSE variant and 1,074 and 1,166 for the VSE variant, respectively.
\begin{table}
\caption{Performance comparison with MATLAB and R}
\label{tab:matlab_r}
\begin{tabular}{lllllllll}
\hline
\multirow{2}{*}{Dataset}&	\multicolumn{2}{c}{Time (s)}&	\multicolumn{3}{c}{Speedup over MLAB}&	\multicolumn{3}{c}{Speedup over R}\\
\hhline{~--------}
&	MLAB&		R&	Na\"{i}ve&	GSE&	VSE&	Na\"{i}ve&	GSE&	VSE\\
\hline
DS21050 & 182,343 & 266,001 & 456 & 435 & 699 & 665 & 635 & 1,019 \\
DS19784 & 205,181 & 284,215 & 484 & 473 & 772 & 671 & 655 & 1,069 \\
DS13070 & 205,063 & 287,635 & 473 & 446 & 743 & 664 & 626 & 1,042 \\
DS3526  & 240,778 & 345,668 & 476 & 494 & 812 & 683 & 709 & 1,166	\\
\hline
\end{tabular}
\end{table}
\subsection{Parallel Scalability Assessment}
\subsubsection{Multithreading}
\begin{figure}
\centering
\begin{subfigure}[b]{0.325\linewidth}
\includegraphics[width=\linewidth]{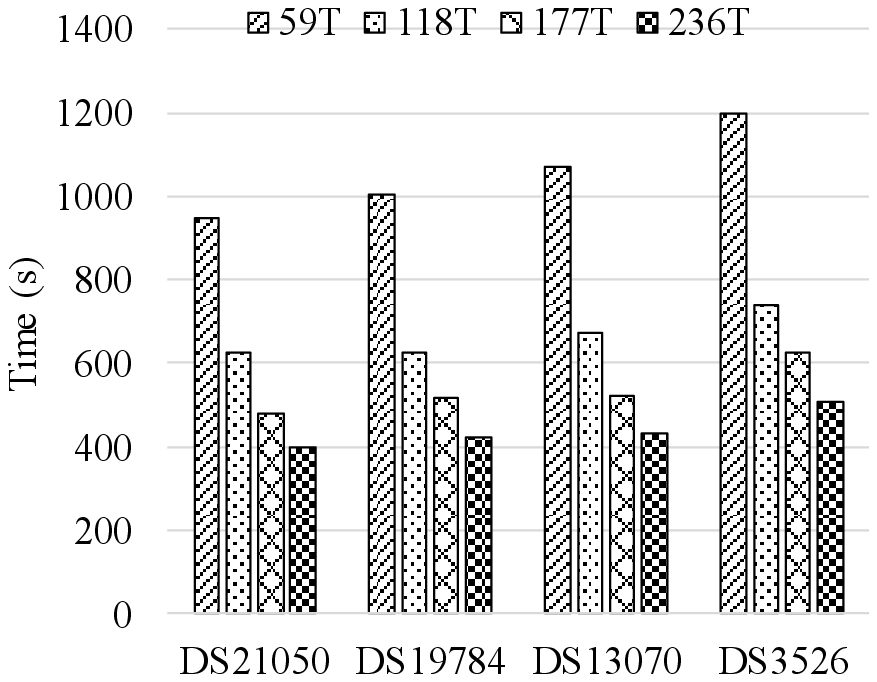}
\caption{}
\label{fig:phi_mt_naive}
\end{subfigure}
\begin{subfigure}[b]{0.325\linewidth}
\centering
\includegraphics[width=\linewidth]{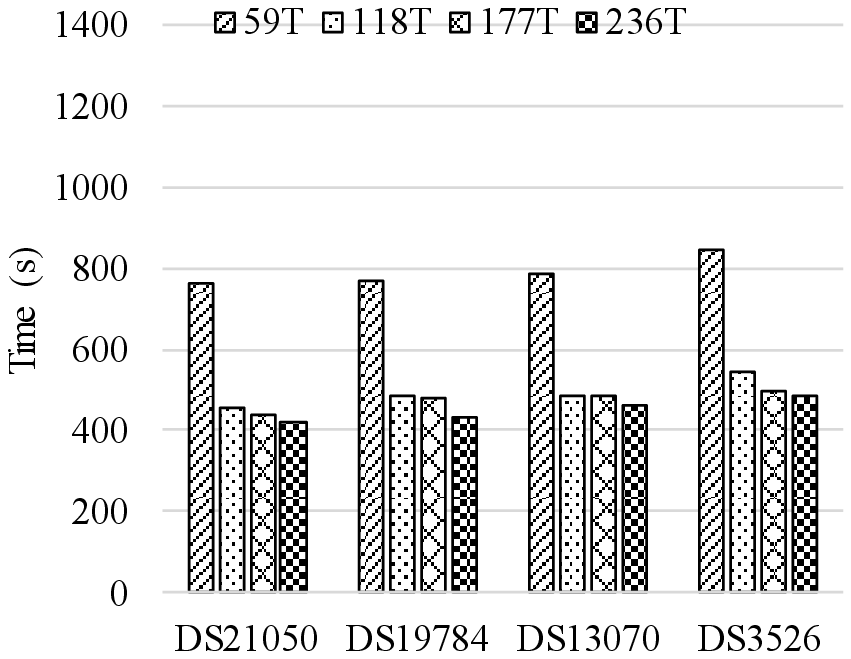}
\caption{}
\label{fig:phi_mt_gse}
\end{subfigure}
\begin{subfigure}[b]{0.325\linewidth}
\centering
\includegraphics[width=\linewidth]{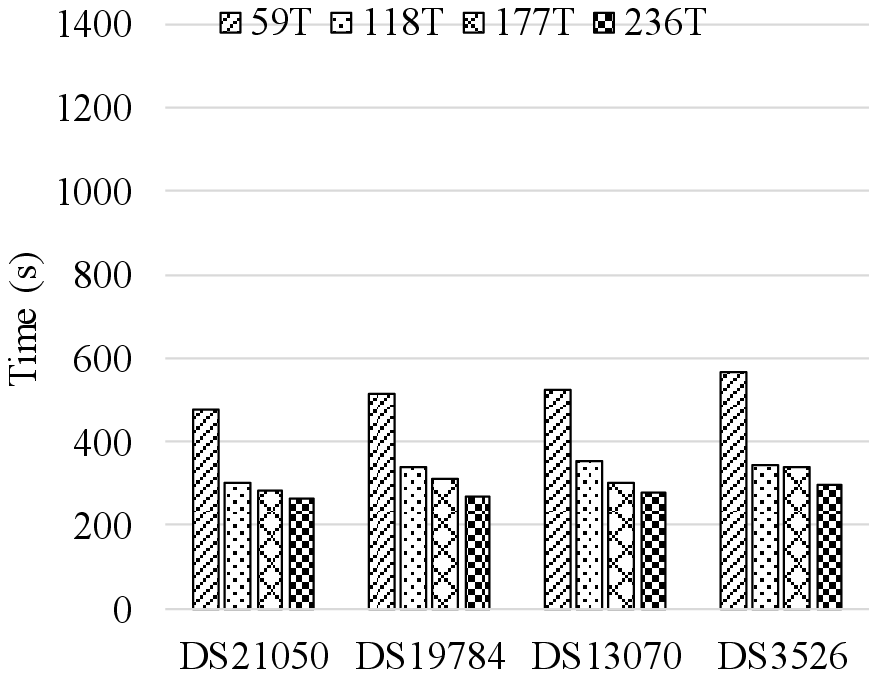}
\caption{}
\label{fig:phi_mt_vse}
\end{subfigure}
\caption{Multithreading scalability of different variants: (a) the na\"{i}ve variant, (b) the GSE variant and (c) the VSE variant.}
\label{fig:mt_scale}
\end{figure}

Thirdly, we evaluated the multithreading scalability of our variants on the Phi with respect to number of threads. Figure \ref{fig:mt_scale} shows the parallel scalability of the three variants. In this test, we used {\tt balanced} thread affinity to ensure that thread allocation is balanced over the cores and the threads allocated to the same core have consecutive identifiers (i.e. neighbors of each other). This thread affinity configuration is attained by setting the environment variable {\tt KMP\_AFFINITY} to {\tt balanced}.

From the figures, it is observed that each variant gets performance improved as the number of active threads per core grows from 1, via 2 and 3, to 4 (corresponding to 59, 118, 177 and 236 threads, respectively). As each core is dual issue and employs in-order execution, at least two threads per core are required in order to saturate the compute capacity of each core. Meanwhile, when moving from one thread per core to two threads per core, we expect that the speedup could be close to 2. In this regard, we investigated the speedup of the instance with two threads per core over the one with one thread per core for each variant, and found that the average speedups are 1.58, 1.61 and 1.56 for the na\"{i}ve, GSE and VSE variants, respectively. This finding is close to our expectation largely. Furthermore, since all threads per core share the dual in-order execution pipeline, deploying three or four threads per core may further improve performance but normally with decreased parallel efficiency (with respect to threads rather than cores). Note that for the 5110P Phi, only 59 out of 60 cores are used for computing as one core is reserved for the operating system running inside. Since each variant reaches peak performance at 236 threads, we used this number of threads for performance evaluation in all tests, unless otherwise stated.
\subsubsection{Distributed computing}
\begin{figure}
\centering
\begin{subfigure}[b]{0.49\linewidth}
\includegraphics[width=\linewidth]{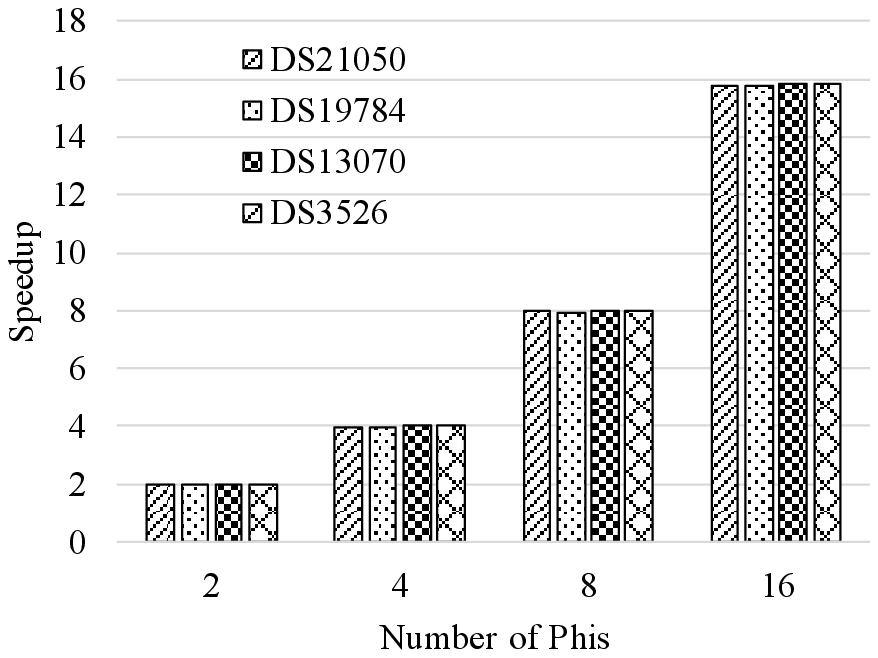}
\caption{}
\label{fig:phi_mpi_naive}
\end{subfigure}
\begin{subfigure}[b]{0.49\linewidth}
\centering
\includegraphics[width=\linewidth]{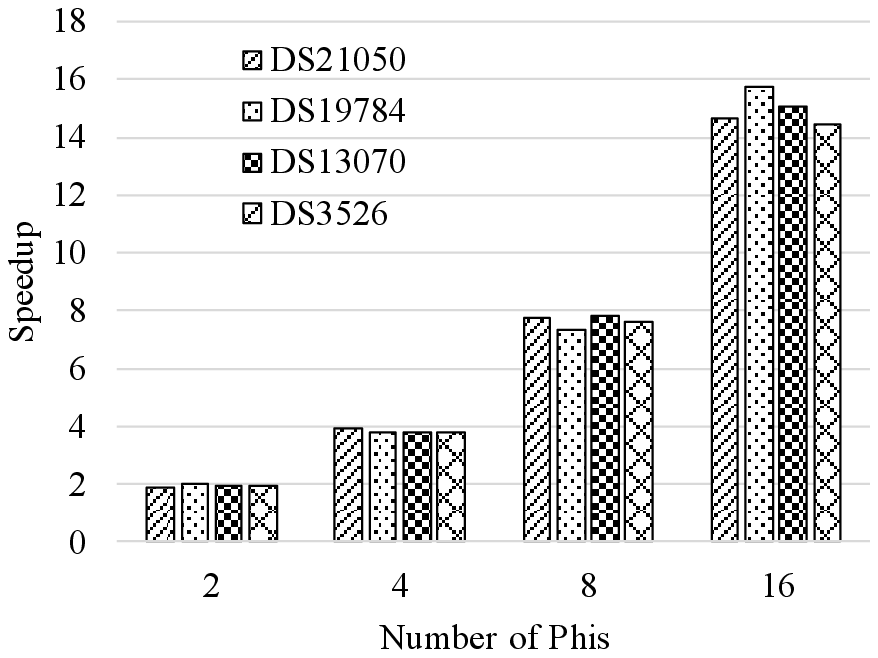}
\caption{}
\label{fig:phi_mpi_gse}
\end{subfigure}
\begin{subfigure}[b]{0.49\linewidth}
\centering
\includegraphics[width=\linewidth]{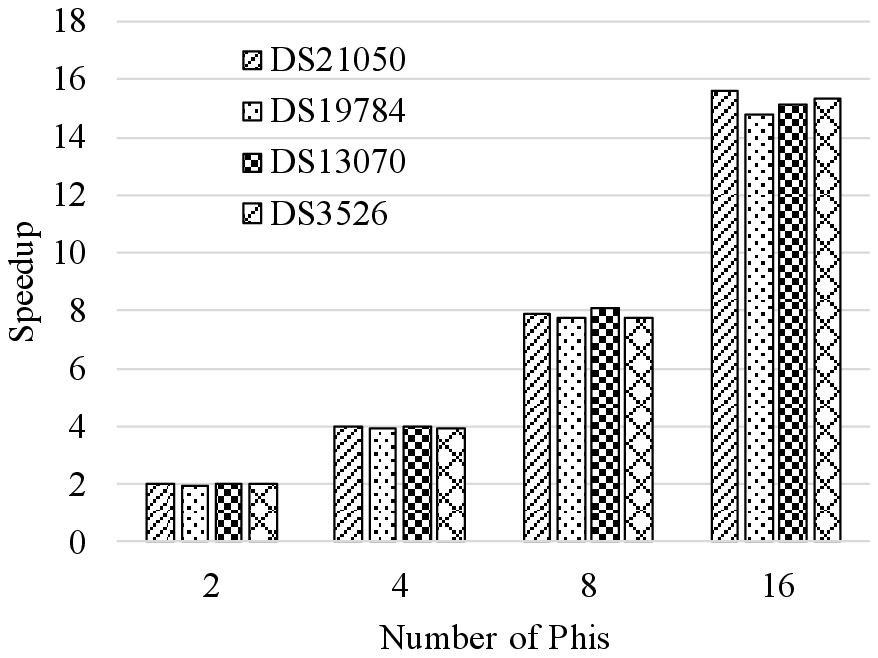}
\caption{}
\label{fig:phi_mpi_vse}
\end{subfigure}
\begin{subfigure}[b]{0.49\linewidth}
\centering
\includegraphics[width=\linewidth]{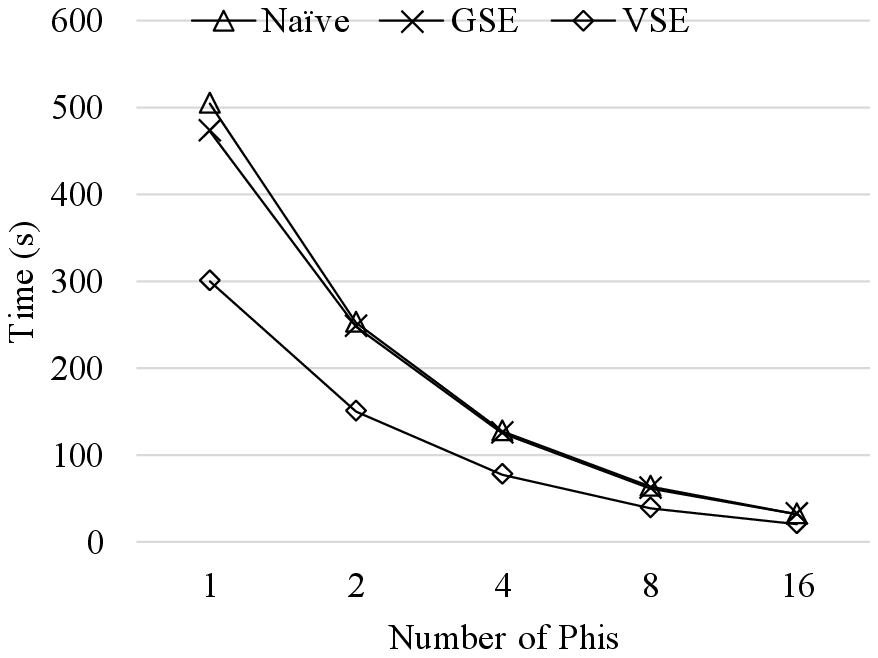}
\caption{}
\label{fig:phi_mpi_ds3526}
\end{subfigure}
\caption{Distributed computing scalability: (a) the na\"{i}ve variant, (b) the GSE variant and (c) the VSE variant and (d) runtimes of each variant on the DS3526 dataset.}
\label{fig:mpi_scale}
\end{figure}

Finally, we measured the parallel scalability of our algorithm by varying the number of Phis used in a distributed environment (see Figure \ref{fig:mpi_scale}). This test is conducted in a cluster that consists of 16 Phis and is constituted by the aforementioned 8 compute nodes. Each variant used 236 threads, since this setting leads to the best performance as mentioned above. From Figure \ref{fig:mpi_scale}, the na\"{i}ve variant demonstrates nearly constant speedups for all datasets on a specific number of Phis, while the two sorting-enabled variants exposed slight fluctuations under the same hardware configuration. This can be explained by the fact that the runtime of our na\"{i}ve variant is independent of vector content, whereas that of each sorting-enabled variant is sensitive to actual vector content to some degree. Moreover, for each dataset, it is observed that every variant has demonstrated rather good scalability with respect to number of Phis. Concretely, the na\"{i}ve variant achieves an average speedup of 2.00, 3.99, 7.96 and 15.81, the GSE one yields an average speedup of 1.93, 3.82, 7.63 and 14.97, and the VSE one produces an average speedup of 1.99, 3.94, 7.86 and 15.23, on 2, 4, 8 and 16 Phis, respectively.
\section{Conclusion}
Pairwise association measure is an important operation in searching for meaningful insights within a dataset by examining potentially interesting relationships between data variables of the dataset. However, all-pairs association computation are computationally demanding for a large volume of data and may take a prohibitively long sequential runtime. This computational challenge for big data motivated us to use parallel/high-performance computing architectures to accelerate its execution.

In this paper, we have investigated the parallelization of all-pairs Kendall's $\tau$ coefficient computation using MIC processors as the accelerator. To the best of our knowledge, this is for the first time that all-pairs $\tau$ coefficient computation has been parallelized on MIC processors. For the $\tau$ coefficient, we have developed three variants, namely the na\"{i}ve variant,  the GSE variant and the VSE variant, based on three pairwise $\tau$ coefficient kernels, i.e. the na\"{i}ve kernel, the GSE kernel and the VSE kernel. The three variants all can execute on CPUs, Phis and their clusters. The na\"{i}ve kernel has a time complexity of $O(n^2)$, while the other two sorting-enabled kernels have an improved time complexity of $O(n\log n)$. Furthermore, we have proposed a generic framework for workload balancing in symmetric all-pairs computation and overcoming memory limitation on the host or accelerator side. This framework assigns unique identifiers to jobs in the upper triangle of the job matrix and builds provable bijective functions between job identifier and coordinate space.

The performance of the three variants was evaluated using a collection of real gene expression datasets produced from whole human genomes. Our experimental results demonstrated that on one 5110P Phi, the na\"{i}ve variant runs up to $25.67\times$ faster than its execution on a single CPU core, and up to $1.87\times$ faster than on 16 CPU cores. On the same Phi, the GSE and VSE variants run up to $15.77\times$ and $16.00\times$ faster than their executions on a single CPU core, and up to $1.15\times$ and $1.15\times$ faster than on 16 CPU cores, respectively. Meanwhile, on the same CPU/Phi hardware configuration, the VSE variant achieves superior performance to the na\"{i}ve and GSE variants. Interestingly, the na\"{i}ve variant was observed to be inferior to the GSE variant on CPUs for both single-threaded and 16-threaded settings, but became superior to the latter on the 5110P Phi for most benchmarking datasets used. As both variants use the same core C++ code for their corresponding CPU- and MIC-oriented implementations as mentioned above, this observation suggests that the architectural differences between different types of PUs may make substantial impact on the resulting performance. Subsequently, we compared our algorithm with the third-party counterparts implemented in the popular MATLAB and R, observing that our algorithm on a single 5110P Phi runs up to $812\times$ faster than 16-threaded MATLAB and up to $1,166\times$ faster than sequential R, both of which are executed on high-end Intel E5-2650 CPUs. We further assessed the parallel scalability of our algorithm with respect to number of Phis in a distributed environment, exposing that our algorithm demonstrated rather good distributed computing scalability.

Finally, it is worth mentioning that the current version of our VSE kernel constrains the largest $n$ value to $2^{15}-1$, due to 32-bit integer packing. One solution to overcome this constraint is packing a variable pair ($u_i$, $v_i$) into a 64-bit integer with each variable occupying 32 bits. In this case, we can split a 512-bit vector into 8 lanes with each lane representing a 64-bit integer, and then implement vectorized pairwise merge of sorted subarrays based on 64-bit integers. Nonetheless, it should be noted that this 64-bit solution has twice less parallelism than 32-bit. In the future, we plan to combine this non-linear measure with other linear or non-linear correlation/dependence measures to generate fused co-expression networks for genome-wide gene expression data analysis at population level. In addition, since R programming language is frequently used in data analytics and the R implementation of all-pairs $\tau$ coefficient computation is extremely slow for large-scale datasets, we also plan to release a R package for public use based on our research in this study.
\section*{Acknowledgment}
This research is supported in part by US National Science Foundation under IIS-1416259 and an Intel Parallel Computing Center award. \textit{Conflict of interest}: none declared.
\section*{References}

\begin{thebibliography}{10}
\expandafter\ifx\csname url\endcsname\relax
  \def\url#1{\texttt{#1}}\fi
\expandafter\ifx\csname urlprefix\endcsname\relax\def\urlprefix{URL }\fi
\expandafter\ifx\csname href\endcsname\relax
  \def\href#1#2{#2} \def\path#1{#1}\fi

\bibitem{zhu2015targeted}
Q.~Zhu, A.~K. Wong, A.~Krishnan, M.~R. Aure, A.~Tadych, R.~Zhang, D.~C. Corney,
  C.~S. Greene, L.~A. Bongo, V.~N. Kristensen, et~al., Targeted exploration and
  analysis of large cross-platform human transcriptomic compendia, Nature
  Methods 12~(3) (2015) 211--214.

\bibitem{steuer2002mutual}
R.~Steuer, J.~Kurths, C.~O. Daub, J.~Weise, J.~Selbig, The mutual information:
  detecting and evaluating dependencies between variables, Bioinformatics
  18~(suppl 2) (2002) S231--S240.

\bibitem{butte1999unsupervised}
A.~J. Butte, I.~S. Kohane, Unsupervised knowledge discovery in medical
  databases using relevance networks., in: Proceedings of the AMIA Symposium,
  American Medical Informatics Association, 1999, p. 711.

\bibitem{mutwil2011planet}
M.~Mutwil, S.~Klie, T.~Tohge, F.~M. Giorgi, O.~Wilkins, M.~M. Campbell, A.~R.
  Fernie, B.~Usadel, Z.~Nikoloski, S.~Persson, Planet: combined sequence and
  expression comparisons across plant networks derived from seven species, The
  Plant Cell 23~(3) (2011) 895--910.

\bibitem{gobbi2015null}
A.~Gobbi, G.~Jurman, A null model for pearson coexpression networks, PloS One
  10~(6) (2015) e0128115.

\bibitem{margolin2006aracne}
A.~A. Margolin, I.~Nemenman, K.~Basso, C.~Wiggins, G.~Stolovitzky, R.~D.
  Favera, A.~Califano, {ARACNE}: an algorithm for the reconstruction of gene
  regulatory networks in a mammalian cellular context, BMC Bioinformatics
  7~(Suppl 1) (2006) S7.

\bibitem{aluru2012reverse}
M.~Aluru, J.~Zola, D.~Nettleton, S.~Aluru, Reverse engineering and analysis of
  large genome-scale gene networks, Nucleic Acids Research 41~(1) (2013) e24.

\bibitem{lachmann2016aracne}
A.~Lachmann, F.~M. Giorgi, G.~Lopez, A.~Califano, Aracne-ap: gene network
  reverse engineering through adaptive partitioning inference of mutual
  information, Bioinformatics 32~(14) (2016) 2233--2235.

\bibitem{lee1988thirteen}
J.~Lee~Rodgers, W.~A. Nicewander, Thirteen ways to look at the correlation
  coefficient, The American Statistician 42~(1) (1988) 59--66.

\bibitem{song2012comparison}
L.~Song, P.~Langfelder, S.~Horvath, Comparison of co-expression measures:
  mutual information, correlation, and model based indices, BMC Bioinformatics
  13~(1) (2012) 328.

\bibitem{spearman1904spearman}
C.~Spearman, Spearman’s rank correlation coefficient, Amer J Psychol 15
  (1904) 72--101.

\bibitem{kendall1948rank}
M.~G. Kendall, Rank correlation methods, Griffin, 1948.

\bibitem{wang2015efficient}
Y.~Wang, Y.~Li, H.~Cao, M.~Xiong, Y.~Y. Shugart, L.~Jin, Efficient test for
  nonlinear dependence of two continuous variables, BMC Bioinformatics 16~(1)
  (2015) 1.

\bibitem{xu2013comparative}
W.~Xu, Y.~Hou, Y.~Hung, Y.~Zou, A comparative analysis of spearman's rho and
  kendall's tau in normal and contaminated normal models, Signal Processing
  93~(1) (2013) 261--276.

\bibitem{darbellay1999estimation}
G.~A. Darbellay, I.~Vajda, et~al., Estimation of the information by an adaptive
  partitioning of the observation space, IEEE Transactions on Information
  Theory 45~(4) (1999) 1315--1321.

\bibitem{daub2004estimating}
C.~O. Daub, R.~Steuer, J.~Selbig, S.~Kloska, Estimating mutual information
  using b-spline functions--an improved similarity measure for analysing gene
  expression data, BMC Bioinformatics 5~(1) (2004) 1.

\bibitem{liang2008gene}
K.-C. Liang, X.~Wang, Gene regulatory network reconstruction using conditional
  mutual information, EURASIP Journal on Bioinformatics and Systems Biology
  2008~(1) (2008) 1--14.

\bibitem{szekely2007measuring}
G.~J. Sz{\'e}kely, M.~L. Rizzo, N.~K. Bakirov, et~al., Measuring and testing
  dependence by correlation of distances, The Annals of Statistics 35~(6)
  (2007) 2769--2794.

\bibitem{kosorok2009brownian}
M.~R. Kosorok, On brownian distance covariance and high dimensional data, The
  Annals of Applied Statistics 3~(4) (2009) 1266.

\bibitem{gretton2005measuring}
A.~Gretton, O.~Bousquet, A.~Smola, B.~Sch{\"o}lkopf, Measuring statistical
  dependence with hilbert-schmidt norms, in: International Conference on
  Algorithmic Learning Theory, Springer, 2005, pp. 63--77.

\bibitem{reshef2011detecting}
D.~N. Reshef, Y.~A. Reshef, H.~K. Finucane, S.~R. Grossman, G.~McVean, P.~J.
  Turnbaugh, E.~S. Lander, M.~Mitzenmacher, P.~C. Sabeti, Detecting novel
  associations in large data sets, Science 334~(6062) (2011) 1518--1524.

\bibitem{song2012feature}
L.~Song, A.~Smola, A.~Gretton, J.~Bedo, K.~Borgwardt, Feature selection via
  dependence maximization, Journal of Machine Learning Research 13~(May) (2012)
  1393--1434.

\bibitem{chang2016gsa}
B.~H.~W. Chang, W.~Tian, Gsa-lightning: ultra-fast permutation-based gene set
  analysis, Bioinformatics 32~(19) (2016) 3029--3031.

\bibitem{abdi2007kendall}
H.~Abdi, The kendall rank correlation coefficient, Encyclopedia of Measurement
  and Statistics. Sage, Thousand Oaks, CA (2007) 508--510.

\bibitem{wang2014optimising}
S.~Wang, I.~Pandis, D.~Johnson, I.~Emam, F.~Guitton, A.~Oehmichen, Y.~Guo,
  Optimising parallel {R} correlation matrix calculations on gene expression
  data using mapreduce, BMC Bioinformatics 15~(1) (2014) 1.

\bibitem{team2013r}
R.~C. Team, et~al., R: A language and environment for statistical computing, R
  Foundation for Statistical Computing, 2013.

\bibitem{dean2008mapreduce}
J.~Dean, S.~Ghemawat, Mapreduce: simplified data processing on large clusters,
  Communications of the ACM 51~(1) (2008) 107--113.

\bibitem{liu2016parallel}
Y.~Liu, T.~Pan, S.~Aluru, Parallel pairwise correlation computation on intel
  xeon phi clusters, in: 28th International Symposium on Computer Architecture
  and High Performance Computing, IEEE, 2016, pp. 141--149.

\bibitem{matlab2015b}
Mathworks, https://www.mathworks.com/products/matlab.html (2015).

\bibitem{sodani2016knights}
A.~Sodani, R.~Gramunt, J.~Corbal, H.-S. Kim, K.~Vinod, S.~Chinthamani,
  S.~Hutsell, R.~Agarwal, Y.-C. Liu, Knights landing: Second-generation intel
  xeon phi product, IEEE Micro 36~(2) (2016) 34--46.

\bibitem{jeffers2013intel}
J.~Jeffers, J.~Reinders, Intel Xeon Phi coprocessor high-performance
  programming, Morgan Kaufmann, 2013.

\bibitem{knight1966computer}
W.~R. Knight, A computer method for calculating kendall's tau with ungrouped
  data, Journal of the American Statistical Association 61~(314) (1966)
  436--439.

\bibitem{inoue2007aa}
H.~Inoue, T.~Moriyama, H.~Komatsu, T.~Nakatani, Aa-sort: A new parallel sorting
  algorithm for multi-core simd processors, in: Proceedings of the 16th
  International Conference on Parallel Architecture and Compilation Techniques,
  IEEE Computer Society, 2007, pp. 189--198.

\bibitem{batcher1968sorting}
K.~E. Batcher, Sorting networks and their applications, in: Proceedings of the
  April 30--May 2, 1968, spring joint computer conference, ACM, 1968, pp.
  307--314.

\bibitem{chhugani2008efficient}
J.~Chhugani, A.~D. Nguyen, V.~W. Lee, W.~Macy, M.~Hagog, Y.-K. Chen,
  A.~Baransi, S.~Kumar, P.~Dubey, Efficient implementation of sorting on
  multi-core simd cpu architecture, Proceedings of the VLDB Endowment 1~(2)
  (2008) 1313--1324.

\bibitem{xiaochen2013register}
T.~Xiaochen, K.~Rocki, R.~Suda, Register level sort algorithm on multi-core
  simd processors, in: Proceedings of the 3rd Workshop on Irregular
  Applications: Architectures and Algorithms, ACM, 2013, p.~9.

\bibitem{odeh2012merge}
S.~Odeh, O.~Green, Z.~Mwassi, O.~Shmueli, Y.~Birk, Merge path-cache-efficient
  parallel merge and sort, Tech. rep., Technical report, CCIT Report (2012).

\bibitem{green2012gpu}
O.~Green, R.~McColl, D.~A. Bader, Gpu merge path: a gpu merging algorithm, in:
  Proceedings of the 26th ACM international conference on Supercomputing, ACM,
  2012, pp. 331--340.

\bibitem{phioptimization2016}
Intel, Intel xeon phi processor software optimization guide,
  https://software.intel.com/en-us/articles/intel-xeon-phi-processor-software-optimization-guide
  (2016).

\bibitem{kiefer2010pairwise}
T.~Kiefer, P.~B. Volk, W.~Lehner, Pairwise element computation with
  {MapReduce}, in: Proceedings of the 19th ACM International Symposium on High
  Performance Distributed Computing, ACM, 2010, pp. 826--833.

\bibitem{gropp2014using}
W.~Gropp, T.~Hoefler, R.~Thakur, E.~Lusk, Using advanced MPI: Modern features
  of the message-passing interface, MIT Press, 2014.

\bibitem{el2006upc}
T.~El-Ghazawi, L.~Smith, Upc: unified parallel c, in: Proceedings of the 2006
  ACM/IEEE conference on Supercomputing, ACM, 2006, p.~27.

\bibitem{gonzalez2016parallel}
J.~Gonz{\'a}lez-Dom{\'\i}nguez, Y.~Liu, B.~Schmidt, Parallel and scalable
  short-read alignment on multi-core clusters using upc++, PLoS ONE 11~(1)
  (2016) e0145490.

\end{thebibliography}

\end{document}